\documentclass[12pt,preprint]{aastex}

\slugcomment{Accepted for Publication in the Astrophysical Journal}
\shortauthors{Fitzpatrick et al.}
\shorttitle{The LMC Eclipsing Binary HV~5936}

\begin{document}

\title{Fundamental Properties and Distances of LMC Eclipsing Binaries: IV. HV~5936{\footnote{Based on observations with the NASA/ESA Hubble Space Telescope, obtained at the Space Telescope Science Institute, which is operated by the Association of Universities for Research in Astronomy, Inc. under NASA contract No. NAS5-26555.}}}

\author{E.L.~Fitzpatrick\altaffilmark{1,4},~I.~Ribas\altaffilmark{2,4},~E.F.~Guinan\altaffilmark{1},~F.P.~Maloney\altaffilmark{1},~A.~Claret\altaffilmark{3}}
\altaffiltext{1}{Department of Astronomy \& Astrophysics, Villanova University, Villanova, PA 19085, USA}
\altaffiltext{2}{Departament d'Astronomia i Meteorologia, Universitat de Barcelona, Av. Diagonal, 647, E-08028 Barcelona, Spain}
\altaffiltext{3}{Instituto de Astrof\'{\i}sica de Andaluc\'{\i}a, CSIC, 
Apartado 3004, E-18080 Granada, Spain}
\altaffiltext{4}{Visiting Astronomer, Cerro Tololo Inter-American Observatory, National Optical Astronomy Observatories, which is operated by the Association of Universities for Research in Astronomy, Inc. (AURA) under cooperative agreement with the National Science Foundation.}

\begin{abstract}
We have determined the fundamental properties and distance of a fourth
eclipsing binary system (EB) in the Large Magellanic Cloud, HV~5936
($\sim$B0.5 V + $\sim$B2 III).  As in our previous studies, we combine
``classical'' EB light curve and radial velocity curve analyses with
modeling of the UV-through-optical spectral energy distribution of
HV~5936 to produce a detailed characterization of the system. In this
paper, we also include an analysis of the high-resolution optical
absorption line spectra of the binary components.  We find HV~5936 to
be an Algol-class system, in which the masses of the primary and
secondary stars have evolved via mass transfer to
their current values of 11.6~$M_{\sun}$ and 4.7~$M_{\sun}$,
respectively. The properties of the primary star are indistinguishable
from those of a ``normal'' star of the same current mass. The secondary
is found to be overluminous for its current mass and exhibits a
factor-of-2 enhancement in its surface He abundance. These results are
compatible with ``Case A'' mass exchange occurring during the core
hydrogen burning phase of the current secondary. The distance derived
to the system, $43.2\pm1.8$ kpc, implies a distance of $\sim$44.3 to
the optical center of the LMC.  This is several kpc closer than found
in our analyses of other systems and we suggest that HV~5936 lies
``above'' the LMC disk.  This is supported by the very low interstellar
H~{\sc i} column density and low $E(B-V)$ found for the system --- both
of which are consistent with expected Milky Way foreground material ---
and may be associated with HV~5936's location near the LMC supergiant
shell LMC~4.
\end{abstract}


\section{Introduction}

This is the fourth in a series of papers presenting results from
detailed analyses of B-type eclipsing binary (EB) systems in the Large
Magellanic Cloud (LMC).  Our primary scientific goals are:  1) to
determine an essentially complete description of the stellar properties
of each system, thus providing tests and constraints for stellar
evolution theory; and 2) to measure precise individual distance for
each system, from which the general distance to the LMC can be
derived. Because of its role as a fundamental calibrator for distance
indicators, the LMC's distance is particularly important for
determining the size scale of the Universe, and its current uncertainty
of 10-15\% contributes considerably to the uncertainty in the Hubble
Constant (e.g., Mould et al. 2000).

\begin{figure*}
\plotone{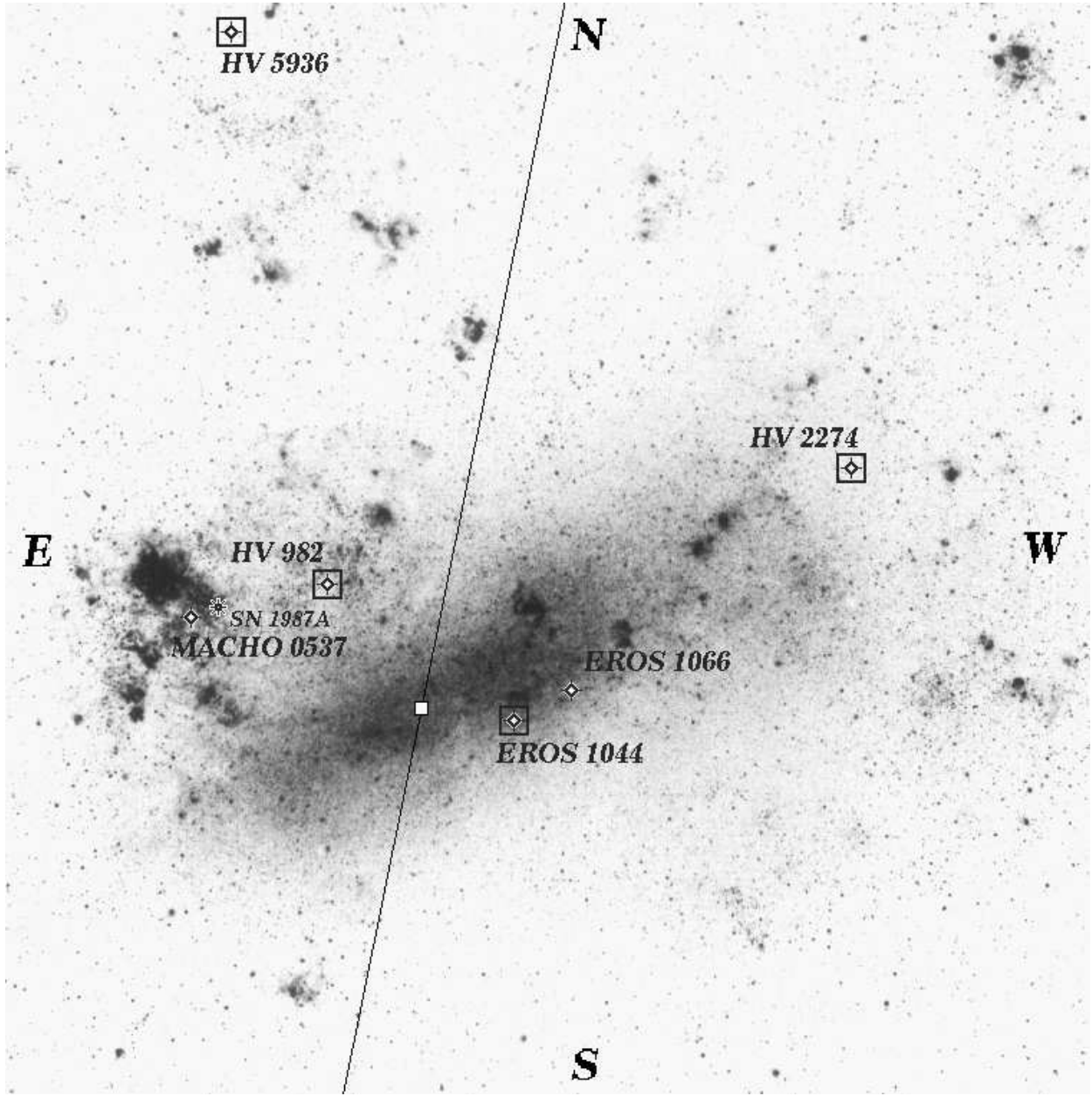}
\caption[hv5936_figLMC.eps]{A photo of the Large Magellanic Cloud
indicating the locations of HV~5936 (this paper), HV~2274 (Paper I),
HV~982 (Paper II), EROS~1044 (Paper III) and two targets of future
analyses, EROS~1066 and MACHO 053648.7-691700 (labeled in the figure as
MACHO 0537).  The optical center of the LMC's bar according to
Isserstedt 1975 is indicated by the open box and the LMC's line of
nodes, according to Schmidt-Kaler \& Gochermann 1992, is shown by the
solid line.  The ``nearside'' of the LMC is to the east of the line of
nodes.  The location of SN 1987A is also indicated. Photo reproduced by
permission of the Carnegie Institution of Washington.
\notetoeditor{THIS FIGURE IS INTENDED TO SPAN TWO COLUMNS}
\label{figLMC}}
\end{figure*}

In our previous studies, we examined the LMC EB systems HV~2274 (Guinan
et al. 1998, hereafter ``Paper I''; Ribas et al. 2000a), HV~982
(Fitzpatrick et al. 2002, ``Paper II''), and EROS~1044 (Ribas et al.
2002, ``Paper III'').  The apparent locations of these systems in the
LMC can be seen in Figure \ref{figLMC}.  The results of these analyses
are beautifully consistent with expectations from stellar structure
theory and provide strong constraints on the distance to the LMC.  As
discussed in Paper III, the three individual distances are all
consistent with a mean of $\sim48$ kpc, although there is a suggestion
that HV~982 and, by association, perhaps the general 30 Doradus region,
may lie behind the LMC's ``Bar'', by several kpc.  Conclusions about
the LMC's distance are somewhat dependent on assumptions about its
spatial orientation and need to be strengthened with additional
measurements from the remaining eclipsing binaries in our program.  Our
approach is ideally suited to pursuing the issues of the spatial
orientation, structure, and general distance to the LMC, since we
measure precise distances to individuals systems which are widely
spread across the face of the LMC.

In this paper, we apply our analysis to a fourth LMC B-type EB,
HV~5936, and derive its stellar properties and distance.  This system,
with $V \simeq 14.8$, stands in some contrast to those in our previous
studies in both its composition and its location. HV~5936 is a
semidetached system, in which the cooler, less massive component fills
completely its Roche lobe. Thus, the currently more massive (and more
luminous) component began its life as the junior, lower mass member of
the binary.  This provides an excellent opportunity to examine the
characteristics of the massive component, which --- due to the rapid
dynamical relaxation expected to follow mass transfer --- should be
indistinguishable from a ``normal'' star of the same current mass.  In
addition to this feature, HV~5936 is located in a very different part
of the LMC from our previous targets (see Fig.  \ref{figLMC}), lying
several degrees north of the LMC's Bar and superimposed on a well known
``hole'' in the LMC's H~{\sc i} distribution (McGee \& Milton 1966),
corresponding to the supergiant shell LMC~4 (Meaburn 1980).  Its
distance should reflect strongly the spatial orientation of the LMC,
i.e., its inclination angle and line of nodes orientation, and could
provide constraints on these factors.

The structure of this paper is similar to our earlier works.  In \S 2
we describe the data included in the study.  In \S 3 we describe, and
present the results from, our standard analysis, which incorporates the
binary's light curve, radial velocity curve, and UV-through-optical
spectral energy distribution.  A study of the high-resolution optical
spectrum of the HV~5936 components is presented in \S 4.  We combine
all our results and give a detailed characterization of the properties
and likely history of the HV~5936 system in \S 5. Some aspects of our
results relating to the interstellar medium towards HV~5936 are
described in \S 6, including an indication of the relative location of
HV~5936 within the LMC. Finally, we derive the distance to the system
and compare it with our previous results in \S 7 and provide some
summary comments in \S 8.

\section{The Data}

Three distinct datasets are required to carry out our analyses of the
LMC EB systems: precise differential photometry (yielding light
curves), medium-resolution spectroscopy (yielding radial velocity
curves), and multiwavelength spectrophotometry (yielding temperature
and reddening information).  Each of these three datasets are described
briefly below. As in our previous papers, the primary ($P$) and
secondary ($S$) components of the HV~5936 system are defined
photometrically and refer to the hotter and cooler components,
respectively.

\subsection{Optical Photometry}

CCD differential photometric observations of HV~5936 were reported by
Jensen, Clausen, \& Gim\'enez (1988). These data were obtained between
1983 and 1984 with a 1.54-m telescope at the European Southern
Observatory (La Silla, Chile).  The published light curves, obtained in
the Johnson $B$ and $V$ passbands, have fairly good phase coverage,
with 44 and 144 measurements, respectively.  According to Jensen et
al., the precision of the individual differential photometric
measurements is better than 0.010~mag. In this study, we adopt the
orbital ephemeris determined by Jensen et al.:
\[\begin{array}{rcrcrc}
T(\mbox{Min I})&=&{\rm HJD}2445657.7911&+&2.8050681 & E \\
\end{array}\]

\subsection{Optical Spectroscopy}

Radial velocity curves for HV 5936, and a number of other LMC EBs, were
derived from optical echelle spectra obtained by us during 6-night and
8-night observing runs in January and December 2000, respectively, with
the Blanco 4-m telescope at Cerro Tololo Inter-American Observatory in
Chile. The seeing conditions during the two runs ranged between 0.7 and
1.8 arcsec. We secured eighteen spectra of HV~5936 -- near orbital
quadratures -- covering the wavelength range 3600--5500~\AA, with a
spectral resolution of $\lambda/\Delta\lambda \simeq 22,000$, and a S/N
of $\sim$20:1. The plate scale of the data is 0.08~\AA~pix$^{-1}$
(5.3~km~s$^{-1}$~pix$^{-1}$) and there are 2.6 pixels per resolution
element. Identical instrumental setups were used for both observing
runs.  The exposure time per spectrum was 1800 sec, sufficiently short
to avoid significant radial velocity shifts during the integrations.
All the HV~5936 observations were bracketed with ThAr comparison
spectra for proper wavelength calibration. The raw images were reduced
using standard NOAO/IRAF tasks (including bias subtraction, flat field
correction, sky-background subtraction, cosmic ray removal, extraction
of the orders, dispersion correction, merging, and continuum
normalization).

\begin{figure*}
\plotone{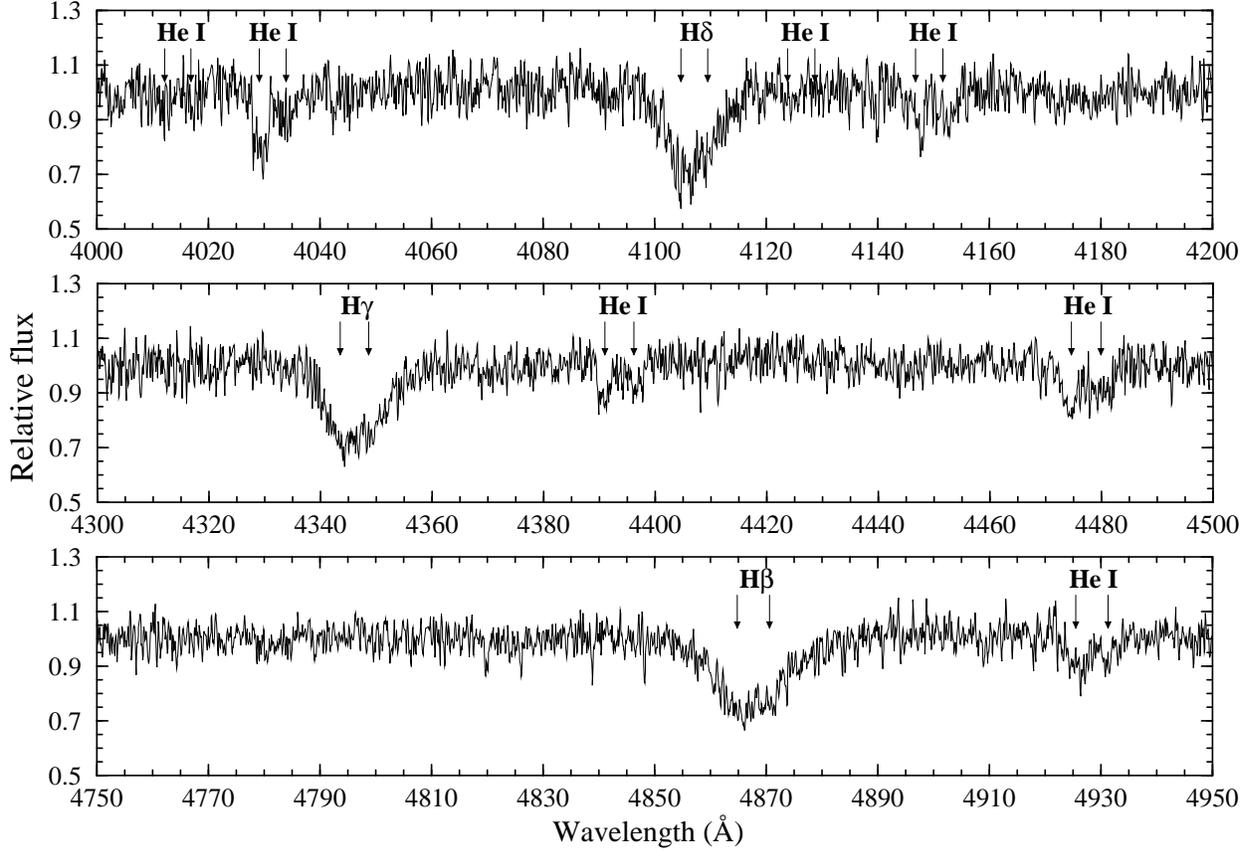} 
\caption[hv5936_figSP.eps]{Normalized
spectrum of HV~5936 near prominent H~{\sc i} and He~{\sc i} lines.  The
spectrum was obtained with the Blanco 4-m telescope at Cerro Tololo
Inter-American Observatory on HJD 2451902.7046 at binary phase 0.297.
The velocity separation between the primary and secondary stars at this
phase is $\Delta$v = 355 km s$^{-1}$.  This figure demonstrates that
the absorption lines from the two stars are cleanly resolved and, thus,
that the radial velocity measurements will be immune to blending
effects.  \notetoeditor{THIS FIGURE IS INTENDED TO SPAN TWO COLUMNS}
\label{figSPEC}}
\end{figure*}

A typical spectrum is shown in Figure \ref{figSPEC}. The H~{\sc i} Balmer
lines and the strongest He~{\sc i} features are labeled, with arrows
marking the expected line positions for the two components of the
system (according to the radial velocity curve solution described in
\S 3.2). This spectrum was obtained at orbital phase 0.30 and
illustrates the clean velocity separation of the two components of the
binary.

To determine the radial velocities of HV~5936's two components from the
echelle spectra, we followed the procedure described in Paper II
utilizing the {\sc korel} program (Hadrava 1995, 1997).  {\sc korel} is
based on the ``spectral disentangling'' technique, which assumes that
an observed double-lined spectrum is a simple linear combination of two
single-lined spectra whose velocities reflect the orbital properties of
the system.  When applied to a set of spectra --- obtained over a range
of orbital phases --- {\sc korel} yields component velocities (relative
to the system barycenter) for each individual spectrum and
``disentangled'' spectra for each of the two binary members, combining
information from the whole ensemble of spectra.  We applied the {\sc
korel} analysis to our echelle data in the 4000--5000~\AA\/ wavelength
region.  The H$\beta$, H$\gamma$ and H$\delta$ lines were excluded by
setting the normalized flux to unity in a window around their central
wavelength.  To translate the velocities to the heliocentric system, we
determined the systemic velocity of HV~5936 from the individual
disentangled spectra.  Cross-correlation of these extracted spectra
with a high S/N ($\sim$250) observed spectrum (HR~1443, B2~IV-V) and
with a synthetic template yielded values consistent with a systemic
velocity of $v_{\gamma}= +314.3\pm5.8$~km~s$^{-1}$.

Table \ref{tabRV} lists the heliocentric radial velocities derived from
all the CTIO spectra using the procedure outlined above (``RV$_P$'' and
``RV$_S$''). Also listed are the dates of observation and the
corresponding phases.  A large number of {\sc korel} runs from
different starting points were carried out to explore the parameter
space and make realistic estimations of the uncertainties. A detailed
discussion of these is left for \S 3.2.

The individual disentangled spectra of the primary and secondary provide
valuable insight into the nature of the HV~5936 system, as well as
confirmation of the results from our general analysis.  These spectra are
discussed in \S 4.

\subsection{UV/Optical Spectrophotometry}

We obtained spectrophotometric observations of HV~5936 at UV and
optical wavelengths with the {\it Hubble Space Telescope} using both
the Faint Object Spectrograph (FOS) and the Space Telescope Imaging
Spectrograph (STIS).  These data are summarized in Table \ref{tabHST}.

The FOS observations utilized the $3.7\arcsec\times1.3\arcsec$
aperture, yielding a spectral resolution of $\lambda/\Delta\lambda
\simeq 1300$.  The four individual spectra were processed and
calibrated using the standard pipeline processing software for the FOS
and then merged  to form a single spectrum which covers the range 1145
\AA\/ to 4790 \AA.  The STIS observations utilized the
$52\arcsec\times0.5\arcsec$ aperture, yielding a spectral resolution of
$\lambda/\Delta\lambda \simeq 750$. These data were processed and
calibrated using the standard pipeline processing software.  Cosmic ray
blemishes were cleaned ``by hand'' and the G430L and G750L spectra were
trimmed to the regions 3510--5690 \AA\/ and 5410--7490 \AA\/,
respectively.  Because of concerns about the photometric zeropoints and
stability, the two STIS spectra were not merged (see \S 3.3).

\section{The Analysis}

Our study of HV~5936 proceeds from three separate but interdependent
analyses. These involve the radial velocity curve, the light curve, and
the spectral energy distribution (SED). The combined results provide
essentially a complete description of the gross physical properties of the
HV~5936 system and a precise measurement of its distance. Each of the
three analyses is described below.
 
\subsection{The Light Curve}

The fits to the light curves were carried out using an improved version
of the Wilson-Devinney (W-D) program (Wilson \& Devinney 1971) that
includes a model atmosphere routine developed by Milone, Stagg \&
Kurucz (1992) for the computation of the stellar radiative parameters.
Both detailed reflection model (MREF=2, NREF=1) and proximity effect
corrections were taken into account when fitting the light curves. The
bolometric albedo and the gravity-brightening coefficients were both
set at the canonical value of 1.0 for stars with radiative envelopes.
For the limb darkening we used a logarithmic law as defined in
Klinglesmith \& Sobieski (1970), with first- and second-order
coefficients interpolated at each iteration for the exact $T_{\rm eff}$
and $\log g$ of each component from a set of tables computed in advance
using a grid of {\sc atlas9} model atmospheres.  A mass ratio of
$q=M_{\rm S}/M_{\rm P}=0.407$ was adopted from the spectroscopic
solution (\S 3.2), and the temperature of the primary component was set
to $T_{\rm eff}^{\rm P} = 26,450$~K, as discussed in \S
\ref{secSED}.  We have adopted a circular orbit as suggested by the
equal width of the eclipses and the occurrence of the secondary eclipse
at exactly phase 0.5.  Further support for this comes from the fact
that the system is semidetached (see below) and orbital circularization
takes place over a very short timescale. Finally, the rotational
velocity of the primary star was set to 1.25 times the synchronous rate
and the secondary star was adopted to rotate synchronously. A
discussion of the component's rotational velocities is provided in \S
4.

Our initial light curve fits were run with a detached configuration
(W-D mode 2). However, preliminary tests indicated a rapid decrease of
the surface gravitational potential of the secondary component until
reaching its critical value. Several runs from different starting
values confirmed this behavior. Therefore, all further W-D solutions
were run in mode 5, i.e., secondary component filling its Roche
critical surface. According to this, HV 5936 is a semidetached binary
where the cooler, less massive component appears to be more evolved and
fills its Roche lobe. This is the classical configuration of post-mass
transfer Algol-class systems. 

In our final analysis we solved for the following light curve
parameters: the orbital inclination ($i$), the temperature of the
secondary ($T_{\rm eff}^{\rm S}$), the gravitational potential of the
primary\footnote{Note that the surface gravitational potential of the
secondary is constrained by the semidetached condition.} ($\Omega_{\rm
P}$), the luminosity of the primary ($L_{\rm P}$), and a phase offset
($\Delta\phi$) which accounts for possible inaccuracy in the adopted ephemeris reference epoch. The iterations with the W-D code were carried out
automatically until convergence, and a solution was defined as the set
of parameters for which the differential corrections suggested by the
program were smaller than the internal probable errors on three
consecutive iterations. As a general rule, several runs with different
starting parameters are used to make realistic estimates of the
uncertainties and to test the uniqueness of the solution.

Because of the relatively small number of free parameters and the
constraint provided by the semidetached configuration, convergence was
achieved rapidly in the light curve fits. The r.m.s. residuals were
determined to be 0.010, and 0.009~mag for the $B$ and $V$ light curves,
respectively. These values are approximately equal to the observed
scatter of the observations. The resulting orbital and physical
parameters are well-defined and the best-fitting model light curves,
together with the O--C residuals, are shown in Figure \ref{figLC}. As
can be seen in the figure, a small systematic departure of the fit
arises in the first quadrature (phase 0.2--0.4) of the $V$ light curve.
This cannot be confirmed in the $B$ light curve because of the lack of
photometric coverage. It is uncertain at this point whether this is a
real effect or just an artifact of the photometric reduction. If it
were real, the asymmetry of the quadrature maxima could arise from the
presence of a hot spot on the primary component. This is not unexpected
in an Algol system because active mass transfer could be taking place
and a hot spot in the atmosphere of the hotter component might arise
from the impact of the accreting material and subsequent kinetic
heating. We explored this scenario by running W-D solutions with an
area on the primary component 10\% hotter than the rest of the
atmosphere. The systematic trend in the residuals did indeed disappear
when the spot was included but this is not surprising because we added
two new free parameters in the analysis (spot radius and location). The
new r.m.s. residuals were found to be 0.009 and 0.008 mag for the $B$
and $V$ light curves, respectively. The solution with the hot spot is
very similar but yields a radius for the primary component about 2\%
smaller and a temperature ratio about 5\% larger.  In the absence of
conclusive evidence, we decided to adopt the ``unperturbed'' solution
without a hot spot.  Besides, the resulting spot location is not
consistent with the expected impact site of a putative stream of matter
using the model of Lubow \& Shu (1975). (Note that the fractional
radius of the primary component is sufficiently large to prevent the
formation of a stable accretion disk (e.g. Albright \& Richards 1993).)

As is evident from Figure \ref{figLC}, the light curves of HV~5936
display rather prominent out-of-eclipse variability. This arises
chiefly from variations in the effective radiating area of the
secondary star.  Due to its non-spherical shape (filling its Roche
lobe), the surface area of the secondary star changes with the orbital
phase and so does its total brightness. The stellar cross-section
reaches a maximum at the orbital quadratures and a minimum at the
eclipses. Another factor contributing to the out-of-eclipse variation
is irradiation. This is a rather significant effect in Algols because
the secondary star is often larger but cooler that the primary. Thus,
when the primary component is in front (near the phase of the secondary
eclipse), we see its light reflected off the atmosphere of the
secondary star. This causes the ingress and egress of the secondary
eclipse to be brighter than the ingress and egress of the primary
eclipse. Both effects discussed here are fully accounted for by the
physical model upon which W-D is based, as proven by the excellent fit
to the light curves.

The final orbital and stellar parameters adopted from the light curve
analysis are listed in Table \ref{tabPARMS}. The uncertainties given in
this table were not adopted from the formal probable errors provided by
the W-D code, but instead from numerical simulations and other
considerations. Several sets of starting parameters were tried in order
to explore the full extent of the parameter space. In addition, the W-D
iterations were not stopped after a solution was found, instead, the
program was kept running to test the stability of the solution and the
geometry of the $\chi^2$ function near the minimum. The scatter in the
resulting parameters from numerous additional solutions yielded
estimated uncertainties that we consider to be more realistic, and are
generally several times larger than the internal statistical errors.

As an internal consistency check, we re-analyzed the light curves with
the mass ratio $q$ left as a free parameter, rather than fixing it to
the spectroscopically determined value.  This test yielded a
``photometric mass ratio'' of $q_{\rm ptm}=0.417\pm0.031$ which is in
excellent with the spectroscopic result of $q=0.407\pm0.016$ (see
\S3.2).  Photometric estimates of q are strongly dependent on
outside-of-eclipse light variations which, in the case of HV 5936,
arise primarily from the changing aspects of the tidally distorted
stars (as well as the reflection effect). The good agreement of $q_{\rm
ptm}$ with the directly determined spectroscopic mass ratio indicates
that the light curves are essentially free of significant perturbing
effects from gas flows and accretion heating --- consistent with our
conclusion above from examining the residuals to the light curve fits.
This agreement reaffirms that the orbital and stellar properties
determined from the combined solutions of light and radial velocity
curves are both self-consistent and robust.

The same light curves as analyzed here (i.e., from Jensen et al. 1988)
have also been studied by Bell et al. (1993). Those authors employed
two different light-curve synthesis programs, one of which was a 1983
version of the W-D code.  Their solutions were run with a mass ratio
quite different from ours ($q_{\rm Bell}=0.46$) and this resulted in a
larger fractional radius for the secondary component than we find.
Also, their adopted temperature for the primary was about 3000~K larger
than our value.  Apart from those, the rest of the light curve
parameters obtained by Bell et al. are compatible with the ones listed
in Table \ref{tabPARMS}. It should be mentioned, however, that our fits
display significantly smaller residuals, which probably result from a
better-determined mass ratio and the more sophisticated fitting program
that we have employed.

\subsection{The Radial Velocity Curve}

\begin{figure*}
\plotone{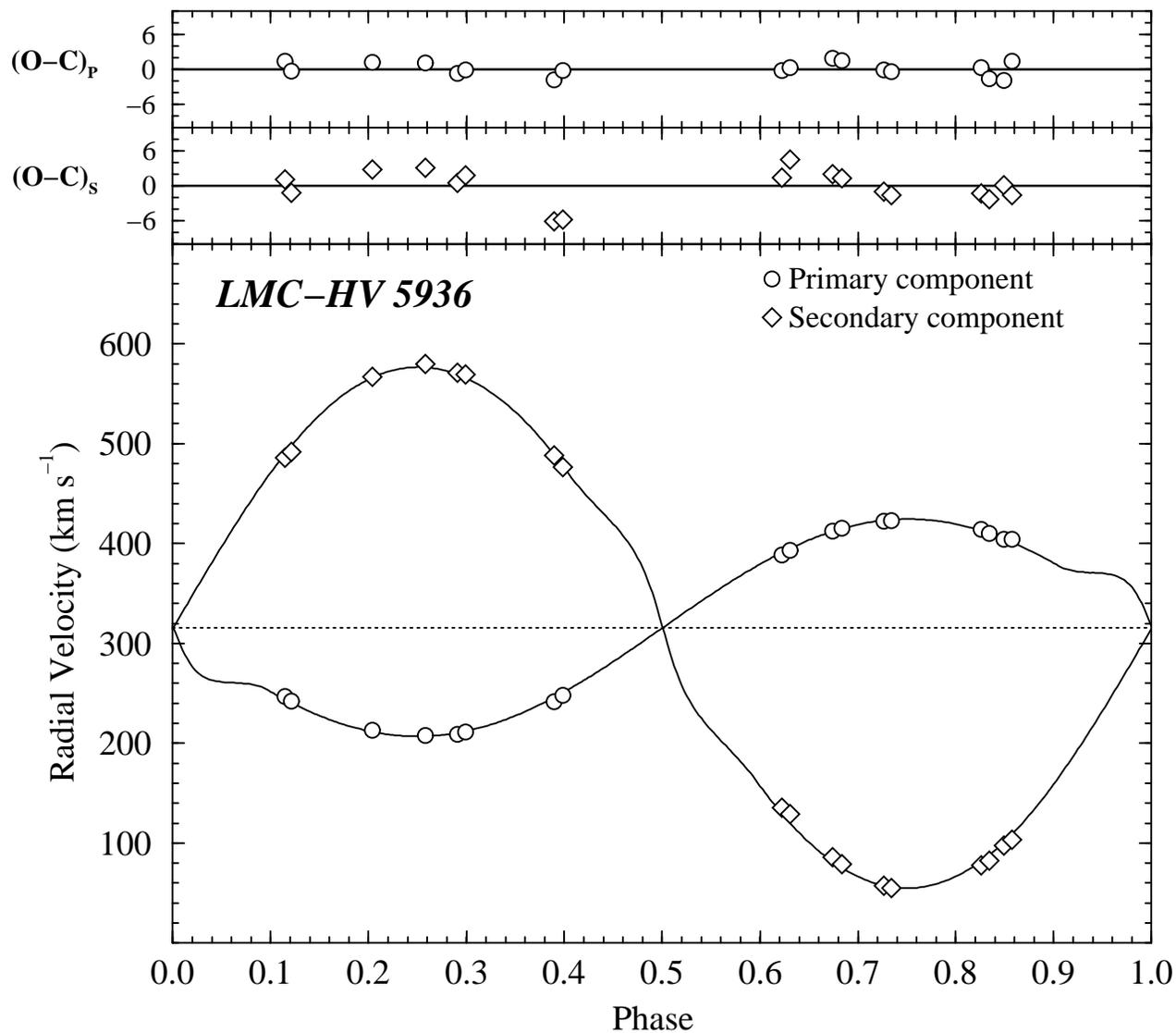}
\caption[hv5936_figRV.eps]{Radial velocity data for HV~5936 (see Table
1) superimposed with best-fitting model. The parameters derived from
the data are listed in Table 2.  Note that the details of the model
curve, including the sharp discontinuity due to the partial eclipse of
a rotating star (the Rossiter Effect), are not a product of the radial
velocity curve analysis. The residuals to the fit are shown in the
upper panel.
\label{figRV}}
\end{figure*}

\begin{figure*}
\plotone{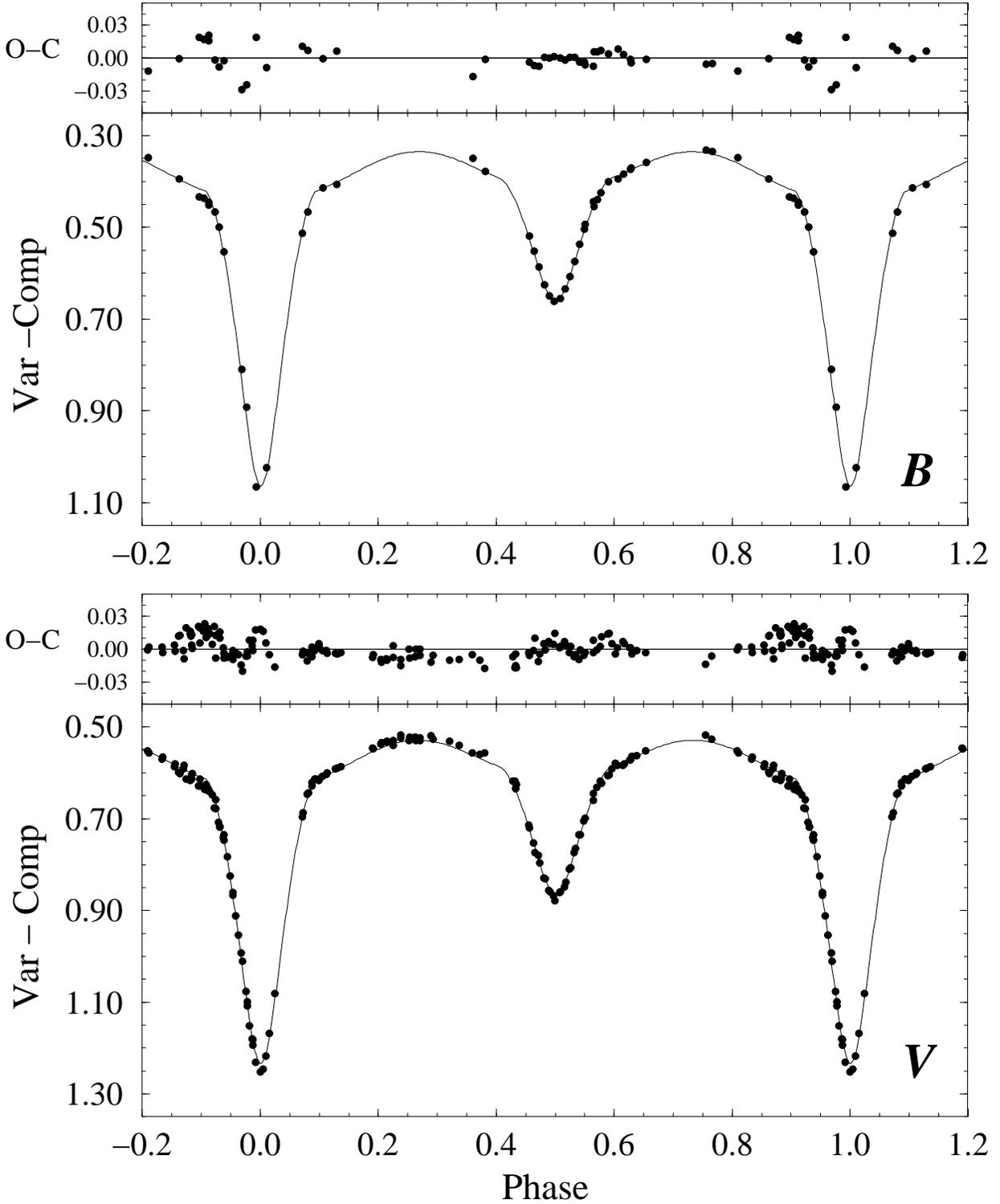}
\caption[hv5936_figLC.eps]{$B$ and $V$ light curves for HV~5936 (filled
circles) overplotted with the best fitting model (solid curves). The
residuals to the fits (``O-C'') are shown above each light curve.  The
parameters derived from the fit are listed in Table 2.
\label{figLC}}
\end{figure*}

The radial velocity curve was fit using the same version of the W-D
program as described above. The free parameters were: the orbital
semi-major axis ($a$), the mass ratio ($q$), and a velocity zero point
(the systemic radial velocity $v_{\gamma}$). The rest of the parameters
were set to those resulting from the light curve solutions discussed in
the previous section. The best fit to the radial velocity curve is
shown in Figure \ref{figRV}. The fit residuals correspond to r.m.s.
internal errors of 1.2 and 2.8~km~s$^{-1}$ for the primary and
secondary components, respectively. The relatively large difference
between these residuals is a consequence of the secondary star being
significantly less luminous, and thus its velocities have intrinsically
larger errors.

The parameters resulting from the radial velocity curve fit are listed
in Table \ref{tabPARMS}. The uncertainties given in the table are not
taken directly from the W-D output, since they fail to account for any
systematic effects that could be present in the velocity data. Instead,
we estimated more realistic errors by considering the scatter of the
velocities derived from the disentangling analysis of separate spectral
regions.  Thus, we divided our entire spectrum into four wavelength
intervals and analyzed these separately with {\sc korel}. The standard
deviation of the resulting velocities was found to be 3.2 and
7.5~km~s$^{-1}$ for the primary and secondary components, respectively.
We conservatively adopted these values as the uncertainty of the
velocity semiamplitudes and scaled the rest of the parameter errors
listed in Table \ref{tabPARMS} accordingly.

Independent radial velocity observations were secured and analyzed by
Bell et al. (1993). According to these authors, the observations were
acquired in less-than-perfect conditions, with atmospheric seeing of
3--6 arcsec and intermittent clouds. Their resulting radial velocity
curves have large scatter, with ``O--C'' residuals up to
80~km~s$^{-1}$. The better quality of our radial velocity data is due
to both the excellent atmospheric conditions at CTIO and the use of the
spectral disentangling technique, which has been proved to be superior
to classical cross-correlation. When comparing the solutions, the value
of the velocity that Bell et al. report for the primary (more luminous)
component is in excellent agreement with our determination listed in
Table \ref{tabPARMS}. However, the velocity semiamplitude of the less
luminous component, more severely affected by poor-quality
observations, found by Bell et al. is much smaller
($\sim$50~km~s$^{-1}$; 3.5$\sigma$) than our value.

\subsection{The UV/Optical Energy Distribution}\label{secSED}
\subsubsection{The Basics}

In general, the observed SED $f_{\lambda\oplus}$ of a binary system can be expressed as:
\begin{eqnarray} \label{basic1}
f_{\lambda\oplus} &=&\left(\frac{R_P}{d} \right)^2 [F_{\lambda}^P + (R_S/R_P)^2 F_{\lambda}^S] \times 10^{-0.4 E(B-V) [k(\lambda-V) + R(V)]}
\end{eqnarray}
where $F_{\lambda}^i$ $\{i=P,S\}$ are the surface fluxes of the primary
and secondary stars, the $R_i$ are the absolute radii of the
components, and $d$ is the distance to the binary. The last term
carries the extinction information, including  $E(B-V)$, the normalized
extinction curve $k(\lambda-V)\equiv E(\lambda-V)/E(B-V)$, and the
ratio of selective-to-total extinction in the $V$ band $R(V) \equiv
A(V)/E(B-V)$.  In our studies, we represent the stellar surface fluxes
with R.L.  Kurucz's {\sc atlas9} atmospheres and use a parameterized
representation of UV-through-IR extinction based on the work of
Fitzpatrick \& Massa (1990) and Fitzpatrick (1999; hereafter F99).  The
Kurucz models are each functions of four parameters ($T_{\rm eff}$,
$\log g$, [m/H], and microturbulence velocity $v_{\rm micro}$), and the
extinction curves are functions of six parameters (see F99). Note that,
for the purposes of Eq. \ref{basic1}, it does not matter which star in
a system is identified as the primary.

We model the observed SED by performing a non-linear least squares fit
to determine the best-fit values of all parameters which contribute to
the right side of equation \ref{basic1}. For HV~5936 (and as for our
previous studies), we can make several simplifications which reduce the
number of free parameters in the problem: (1) the temperature ratio of
the two stars is known from the light curve analysis; (2) the surface
gravities can be determined by combining results from the light and
radial velocity curve analyses and are log g = 3.98 and 3.49 for the
primary and secondary stars, respectively (see \S 5); (3) the values of
[m/H] and $v_{\rm micro}$ can be assumed to be identical for both components; (4)
the ratio $R_{\rm S}/R_{\rm P}$ is known; and (5) the standard mean
value of $R(V) = 3.1$ found for the Milky Way can reasonably be assumed
given the existing LMC measurements (e.g., Koornneef 1982; Morgan \&
Nandy 1982; see \S 6).

We prepared the spectrophotometric datasets for the SED analysis by (1)
velocity-shifting to bring the centroids of the stellar features to
rest velocity; (2) correcting for the presence of a strong interstellar
H~{\sc i} Ly$\alpha$ absorption feature in the FOS spectrum at
1215.7~\AA\/ (see \S 6); and (3) binning to match the {\sc atlas9}
wavelength grid. The statistical errors assigned to each bin were
computed from the statistical errors of the original data, i.e.,
$\sigma_{\rm bin}^2 = 1/\Sigma(1/\sigma_i^2)$, where the $\sigma_i$ are the
statistical errors of the individual spectrophotometric data points
within a bin. For all the spectra, these binned uncertainties typically
lie in the range 0.5\% to 1.5\% of the binned fluxes.  The
weighting factor for each bin in the least squares procedure is given
by $w_{\rm bin} = 1/\sigma_{\rm bin}^{2}$.  We exclude a number of individual
bins from the fit (i.e., set the weight to zero) for the reasons
discussed by FM99 (mainly due to the presence of interstellar gas
absorption features).

As discussed in Paper II, we do not merge the FOS and STIS data into a
single spectrum, but rather perform the fit on the three binned spectra
simultaneously and independently.  We assume that the FOS fluxes
represent the ``true'' flux level and account for zero point
uncertainties in the STIS data by incorporating two zeropoint
corrections (one for each STIS spectrum) in the fitting procedure.  We
later explicitly determine the uncertainties in the results introduced
by zeropoint uncertainties in FOS.

\subsubsection{Special Considerations for HV~5936}

As noted in \S 2.1, the out-of-eclipse variations seen in HV~5936's
light curve result primarily from changes in the apparent size (i.e.,
as presented toward the Earth) of the secondary due to its mild
non-sphericity and also from phase-modulated reflection of light from
the primary off the secondary.  These effects must be taken into
account in the SED analysis because they affect the relative
contributions of primary and secondary light to the observed SED.

We incorporate the effects through a simple modification of Eq. \ref{basic1}:
\begin{eqnarray} \label{basic2}
f_{\lambda\oplus} &=&\left(\frac{R_P}{d} \right)^2 [k_P F_{\lambda}^P + k_S (R_S/R_P)^2 F_{\lambda}^S] \times 10^{-0.4 E(B-V) [k(\lambda-V) + R(V)]}
\end{eqnarray}
where $k_P$ and $k_S$ are phase-dependent correction factors accounting
for additional reflected light from the primary and the varying
apparent size of the secondary, respectively.  The values of $k_P$ and
$k_S$ for each of the spectrophotometric observations are listed in
Table \ref{tabHST} and were computed from the results of the W-D light
curve analysis.  The $k_P$ are always greater than 1, since reflection
always adds otherwise-unseen light from the primary. The value of $k_S$
for the FOS/G130H spectrum indicates that, at the phase of this
observation, the apparent size of the secondary was slightly smaller
than its mean value (as given by the ``volume radius'' computed by the
W-D program).  For all the other spectra, the apparent size of the
secondary was larger than its mean value.  Note that all the
corrections for the FOS datasets --- which provide the photometric
zeropoint for the SED analysis --- are very close to 1.0.
 
\subsubsection{Results}

We computed the final fit to HV~5936's SED utilizing Eq. \ref{basic2}
with the appropriate values of $k_P$ and $k_S$ inserted for each
dataset.  As in previous papers, we adjusted the weights in the fitting
procedure to yield a final value of $\chi^2 = 1$ --- since the
statistical errors of the data under-represent the total uncertainties
(see the discussion in Paper II).  This was accomplished by
quadratically adding an uncertainty equivalent to 1.9\% of the local
binned flux to the statistical uncertainty of each flux point.
(Essentially identical results occur if the statistical errors are
simply scaled by a factor of 2.2 to yield $\chi^2 = 1$.)  This value of
1.9$\%$ gives an indication of the general quality of the fit to
HV~5936's SED, excluding the effect of statistical noise.  It is
comparable the quality level we have seen in the previous analyses.

\begin{figure*}
\plotone{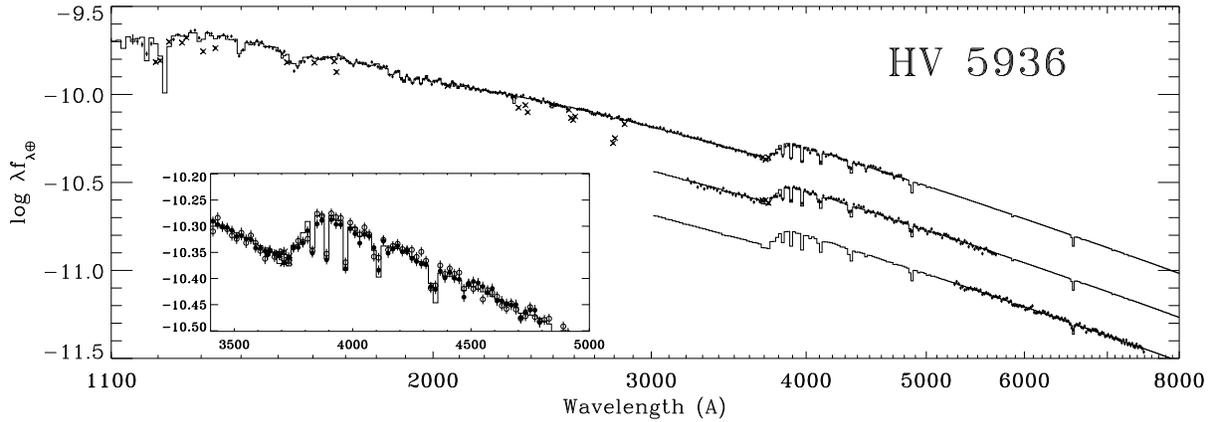}
\caption[hv5936_figSED.eps]{The observed UV/optical energy distribution
of the HV~5936 system (small filled circles), superimposed with the
best-fitting model, consisting of a pair of reddened and
distance-attenuated Kurucz {\sc atlas9} atmosphere models
(histogram-style lines).  Vertical lines through the data points
indicate the $1\sigma$ observational errors.  Crosses denote data
points excluded from the fit, primarily due to contamination by
interstellar absorption lines.  The top spectrum shows the FOS data,
the middle spectrum (shifted by $-0.25$ dex) the STIS/G430L data, and
the lower spectrum (shifted by $-0.5$ dex) the STIS/G750L data.  The
energy distribution fitting procedure was performed simultaneously on
all three datasets.  The inset shows a blowup of the region surrounding
the Balmer Jump which illustrates the overlap between the FOS (solid
circles) and STIS/G430L (open circles) data.  The parameters derived
from the fit to the energy distribution are listed in Tables
\ref{tabPARMS}, \ref{tabSTIS}, and \ref{tabEXT}.  The various
constraints imposed on the fit are discussed in \S 3.3.3.
\notetoeditor{THIS FIGURE IS INTENDED TO SPAN TWO COLUMNS}
\label{figSED}}
\end{figure*}

The best-fitting values of the energy distribution parameters for
HV~5936 and their associated 1$\sigma$ uncertainties (``internal
errors'') are listed in Tables \ref{tabPARMS} (stellar properties),
\ref{tabSTIS} (STIS offsets), and \ref{tabEXT} (extinction curve
parameters).  A comparison between the observed spectra and the
best-fitting model is shown in Figure \ref{figSED}.  The three binned
spectra are plotted separately in the figure for clarity (small filled
circles).  The zeropoint offset corrections (see Table \ref{tabSTIS})
were applied to all STIS spectra in Figure \ref{figSED}.  Note that we
show the quantity $\lambda$f$_{\lambda\oplus}$ as the ordinate in
Figure \ref{figSED} (rather than f$_{\lambda\oplus}$) strictly for
plotting purposes, to ``flatten out'' the energy distributions.

The correction factors of 10.2\% and 5.9\% are required to rectify the
STIS G430L and G750L spectra, respectively, are similar to the results
found in Papers II and III.  This apparently systematic effect probably
results from small light losses in the STIS $0\farcs5$ slit.  This will
be tested by using a wider slit in future STIS observations.

\section{Supplemental Analysis of the ``Disentangled'' Spectra}

In \S 2.2, we utilized the {\sc korel} program primarily to determine
the radial velocities of HV~5936's component stars.  However,
byproducts of this program, i.e., a high-resolution, ``disentangled''
optical spectrum for each star, can provide valuable additional
information on the binary system which is {\it entirely independent} of
the analysis described in the preceding section.  This information can
be tapped by modeling these absorption line spectra with synthetic
spectra.  The potential value of such an analysis is threefold, since
it can: 1) add new information on the system, 2) independently verify
results of the preceding analysis, or 3) identify problems with the
preceding analysis.

\begin{figure*}
\epsscale{0.7}
\plotone{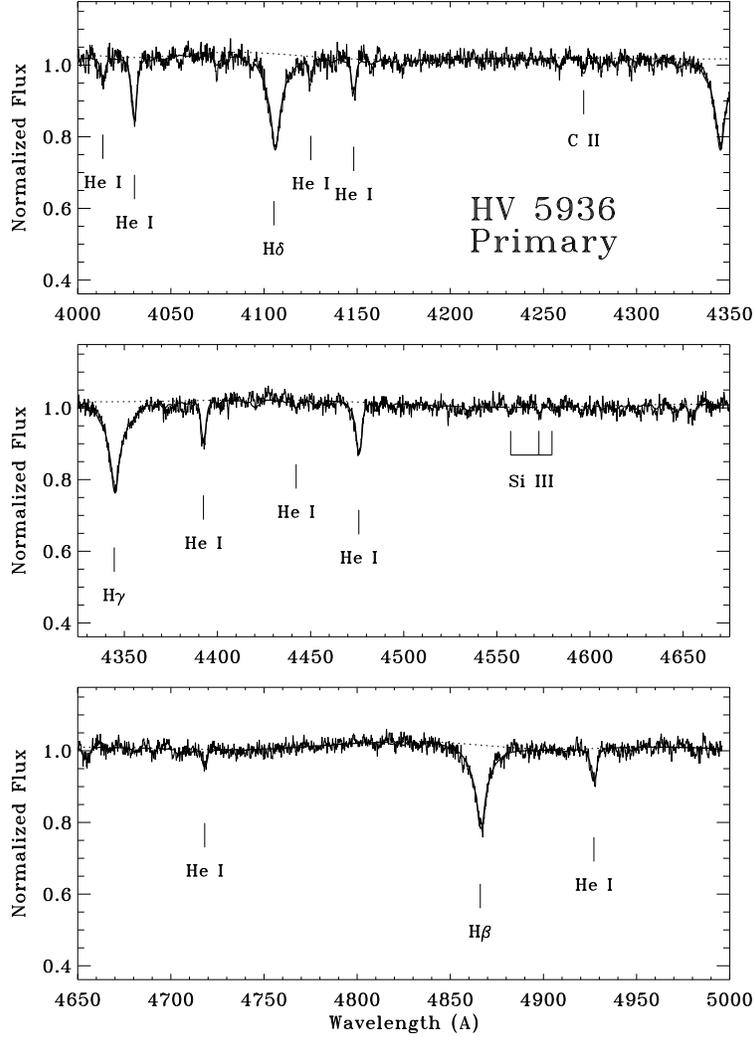}
\caption[hv5936_figPRIM.eps]{``Disentangled'' optical spectrum of the
primary star in the HV~5936 system, as produced by the {\sc korel}
program (see \S 2).  The spectrum consists of 12453 data points, has a
spectral resolution of 0.2 \AA, and has been smoothed by 3 points for
the presentation.  Line strengths in disentangled {\sc korel} spectra
are diluted by the presence of continuum light from the companion
star.  In this case, the companion (i.e., the system's secondary)
contributes about 36.1\% of the continuum light and the y-axis of the
plots have been adjusted so that the flux zeropoint for the primary's
lines corresponds to the bottom of each panel.  The thick solid curve
shows a synthetic spectrum fit to the primary's spectrum, utilizing the
{\sc atlas9} and {\sc synspec} programs.  This fit included the
determination of a 20-order Legendre polynomial to accommodate
undulations in the stellar continuum not removed in the original data
normalization. See Figure \ref{figSPEC} for an example of the original
data. The polynomial itself is shown as the dotted line.  The stellar
properties determined from the fit to the spectrum are consistent with
those derived in \S 3 from the binary and spectral energy distribution
analyses.  See the discussion in \S 5 for more information.
\label{figPRIM}}
\end{figure*}

\begin{figure*}
\epsscale{0.7}
\plotone{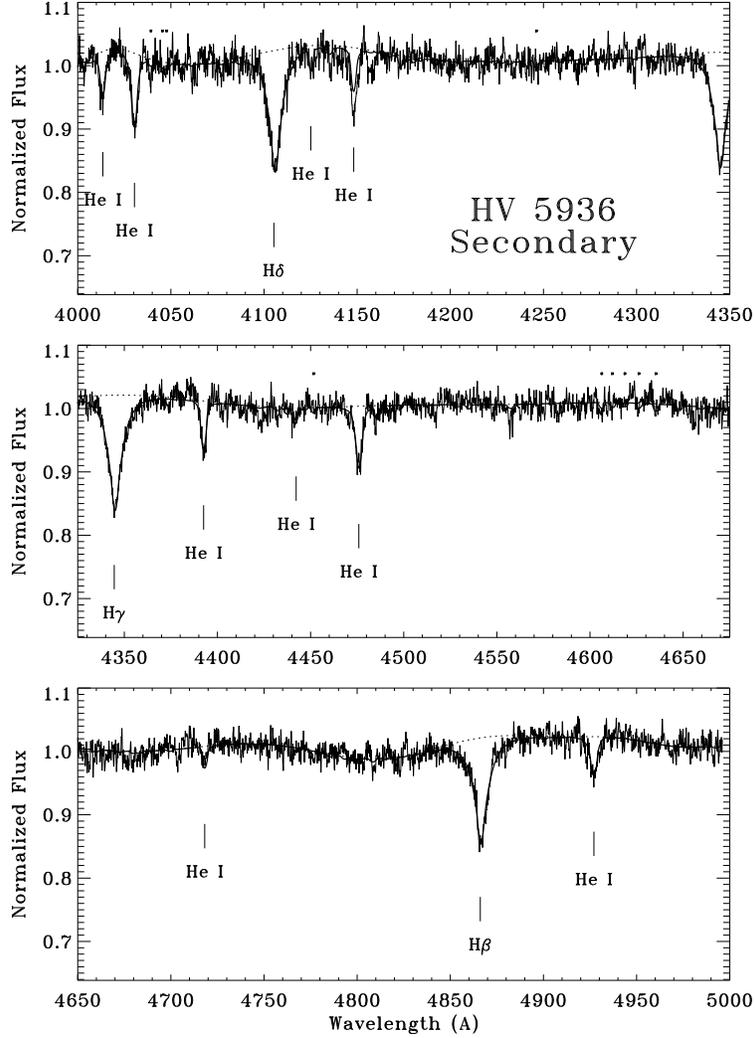}
\caption[hv5936_figSECON.eps]{Same as Fig. \ref{figPRIM} but for the
secondary star in the HV~5936 system. In this case, the secondary's
line strengths are diluted by the presence of the primary, which
contributes about 63.9\% of the continuum light. The flux zeropoint for
the secondary corresponds to the bottom of each panel. The synthetic
spectrum fit (thick solid curve) included the determination of a
25-order Legendre polynomial (dotted line) to accommodate deficiencies
in the original normalization. This fit shown in the figure corresponds
to Case II in Table \ref{tabFINE} and includes an enhanced He abundance
and modified CNO abundances.  It reproduces the stellar properties
derived independently in \S 3. The small asterisks indicate the
locations of the strongest features of N II.  See the discussion in \S
5 for more information.
\label{figSECON}}
\end{figure*}

The disentangled spectra of HV~5936's two components, as produced by
the {\sc korel} program are shown in Figures \ref{figPRIM} and
\ref{figSECON}.  Each spectrum has a resolution of $\sim$0.2 \AA\/ and
consists of 12,453 data points. The most prominent features in both
are lines of H~{\sc i} and He~{\sc i}. The strong C~{\sc ii}
$\lambda$4267 and Si~{\sc iii} $\lambda\lambda$4553,4568,4574 lines are
noted in the primary's spectrum (Fig. \ref{figPRIM}).  Virtually all
the other, weaker absorption features in that spectrum are due to lines
of O~{\sc ii}.  There are no positive identifications of individual
metal absorption lines in the noisier spectrum of the secondary.  The
ripple-like structure in the range 4140--4170 \AA\/ of both spectra
(near the position of a He~{\sc i} line) is an artifact of the {\sc
korel} program, and results because the observations were mainly
obtained near orbital quadratures and not distributed more uniformly
throughout the orbit.  The strengths of all the lines in these two
spectra are diluted by the presence of continuum light from both binary
components.  Given the phase distribution of the original optical data
(see Table 2), we compute that the primary and secondary contribute
63.9\% and 36.1\%, respectively, of the continuum light in these mean
spectra.  These values incorporate the reflection and ellipticity
effects noted in \S 3, and have uncertainties of order $\sim$1\%.  In
Figures \ref{figPRIM} and \ref{figSECON}, we have adjusted the vertical
axes so that the lines are in their correct strengths relative to the
bottoms of each panel.

We model the disentangled spectra by utilizing {\sc atlas9} atmospheric
structure models, Ivan Hubeny's spectral synthesis program {\sc
synspec}, and essentially the same $\chi^2$-minimization technique as
described in \S 3.3 for the SED analysis.  In general, finding the
best-fitting synthetic spectrum for these stars requires the
determination of seven parameters.  Four of these --- $T_{\rm eff}$,
$\log g$, [m/H], and $v_{\rm micro}$ --- are required to specify the
appropriate {\sc atlas9} model.  Two more --- $v_{\rm radial}$ and $v \sin
i$ (along with $v_{\rm micro}$) --- are used explicitly by {\sc
synspec} in determining line positions and widths.  A final parameter,
which we characterize as the percent contribution of each star to the
observed continuum, is required to reproduce the dilution of the line
strengths.  For the case of the HV~5936 stars, we found that the
microturbulence velocity was poorly determined and so we simply adopted
the value of 2.6 km s$^{-1}$ derived in the SED analysis in \S 3.3. In
addition, we simultaneously determined the coefficients of a high-order
Legendre polynomial to allow the smoothing out of ``bumps and wiggles''
which can be seen in the spectra of both stars.  These result from
deficiencies in the normalization of the original echelle spectra used
by {\sc korel}.  The 4140--4170 \AA\/ region, noted above, was excluded
from all fits.  

When analyzing the primary's spectrum we found that the results of the
fitting procedure depended somewhat on the assumed order of the
normalizing function.  Therefore, we performed 7 independent fits,
utilizing 10, 15, 20, 25, 30, 35, or 40 order Legendre polynomials.
These sample the reasonable range, since a polynomial with fewer than
10 orders cannot match the observed structure in the continuum, while
those with more than 40 orders have too much freedom and can distort
the observed features.  The results of these fits are given in column
(2) of Table \ref{tabFINE} where we list the simple means of the
parameters derived from the 7 fits.  The uncertainties quoted in the
table are the quadratic sum of the internal uncertainties in a single
fit and the standard deviation of the ensemble of 7 fits.  In Figure
\ref{figPRIM} we show the best fitting model corresponding to the case
with a 20-order Legendre normalizing function (thick solid curve).  The
polynomial itself is shown by the dotted line.  The results from this
particular fit are very close to the ensemble averages in Table \ref{tabFINE}.

The fit to the primary's disentangled spectrum is very good.  The
results for $T_{\rm eff}$, [m/H], and the contribution factor are all
consistent at the 1$\sigma$ level with previous determinations
(26,450 K, $-0.63$, and 63.9\%, respectively).  The value of $\log g$
is 0.12 dex ($\sim$2.2$\sigma$) higher than that determined from the
binary analysis.  However, this is not a large discrepancy and, since
the gravity determination depends on the wings of the Balmer lines, we
suspect it arises from deficiencies in the normalization of the
original echelle spectra.  It is interesting that the metallicity
determined here --- which mainly reflects weak but numerous lines of
O~{\sc ii} --- agrees so well with that derived by the SED analysis ---
based mainly Fe~{\sc ii} and {\sc iii} absorption in the UV.  These two
results present a very consistent picture of a general factor-of-4
metal underabundance for HV~5936.  Note that all of the fits to the
primary's spectrum assume a helium abundance of $n({\rm He})/n({\rm H})
= 0.084$, which is based on the observed metal abundance and standard
chemical enrichment laws (see \S 5) and corresponds to a helium mass
fraction of $Y=0.25$.

Fits to the secondary's spectrum were performed in a similar manner as
above except that --- since the optical metal lines are all too weak to
allow a meaningful determination of [m/H] --- we fixed it's value to
$[{\rm m/H}] = -0.63$ as derived in \S 3.  Also, we found that the
10-order Legendre fit was inadequate to model the continuum undulations
and so we base our results on 6 fits with normalization polynomials of
15, 20, 25, 30, 35, and 40 orders.  The resulting mean parameters and
1$\sigma$ uncertainties from these fits are listed in column (3) of
Table \ref{tabFINE}.  These results are much less satisfying than for
the primary. $T_{\rm eff}$, $\log g$, and the contribution factor all
differ greatly from the previously determined values (17,600 K, 3.49,
and 36.1\%, respectively).  In addition, the value of $v \sin i$
appears inconsistent with the virtually unavoidable requirement that
the secondary's rotation be tidally locked to its orbit (see \S 5).

From detailed examination of the fits to the secondary, we found that
the discrepancies arise essentially because the secondary's He~{\sc i}
lines are too strong to be well-fit with the temperature of 17,600 K
inferred earlier in \S 3. It is not likely, however, that the secondary
can be significantly hotter than this, since this temperature results
from a very well-determined eclipse-based temperature ratio and an
apparently well-determined primary star temperature.  Rather, we
suggest that the secondary's He~{\sc i} lines are enhanced in strength
arising from a modest enhancement in the He surface abundance.  The viability
of this suggestion is demonstrated by Case II in column (4) of Table
\ref{tabFINE}.  The mean parameters and 1$\sigma$ uncertainties were
derived in an identical manner to those in Case I, except that the
helium abundance was approximately doubled to a value of $n({\rm
He})/n({\rm H}) = 0.16$.  In addition, the abundances of C and O were
depleted by a factor of 10, compared to the base metallicity of [m/H] =
-0.626, and N was enhanced by a factor of 70.  These factors are based
on stellar interior models which will be discussed below in \S 5.  It
is clear that these adjustments to the elemental abundances yield
results completely consistent with the previous measurements of $T_{\rm
eff}$ and $\log g$ for the secondary and the known value of the
continuum contribution factor.  In addition the value of $v \sin i$ is
well-determined and consistent with synchronous rotation (see \S 5).
In Figure \ref{figSECON} we show the best-fitting model corresponding
to Case II with a 25-order Legendre normalizing function (thick solid
curve).  The results from this particular fit are very close to the ensemble
averages in column (4) Table \ref{tabFINE}.

The value of $n({\rm He})/n({\rm H})$ used in Case II was not arrived
at arbitrarily.  Rather, we ran a set of fits to the secondary's
spectrum (each with a different order Legendre polynomial) with $T_{\rm
eff}$ held fixed at the value of 17,600 K found in \S 3 and with the
He abundance as a free parameter.  ($T_{\rm eff}$ and $n({\rm
He})/n({\rm H})$ are degenerate in their effects and cannot both be
determined from fitting the spectrum.)  This set of models resulted in
a best-fitting He abundance of $n({\rm He})/n({\rm H}) = 0.157 \pm
0.018$, which we rounded off to use in Case II.  The significance of
this result as an indicator of a true He enhancement can be gauged from
a similar analysis of the primary. A set of fits of the primary's
spectrum, with $T_{\rm eff}$ fixed at 26,450 K (from the SED analysis)
and the He abundance as a free parameter, yielded a best fitting He
abundance of $n({\rm He})/n({\rm H}) = 0.090 \pm 0.007$.  This is
consistent with that inferred in \S 5 below, from the metallicity of
the system and chemical enrichment laws.

Our conclusions from analysis of the high-resolution disentangled
optical spectra of HV~5936 are twofold:  1) the primary's absorption
line spectrum is completely consistent with the stellar properties
derived in \S 3, given the random errors and small systematic effects
which may have arisen during the processing and ``disentangling'' of
the spectra; and 2) the secondary's spectrum is consistent with the
derived properties only if a $\sim$factor-of-2 enhancement is assumed
in the surface He abundance, as compared to that of the primary.  In \S
5 below we consider the plausibility of a He enhancement in the
secondary.

\section{The History and Nature of the HV~5936 Binary System}

\begin{figure*}
\epsscale{1.0}
\plotone{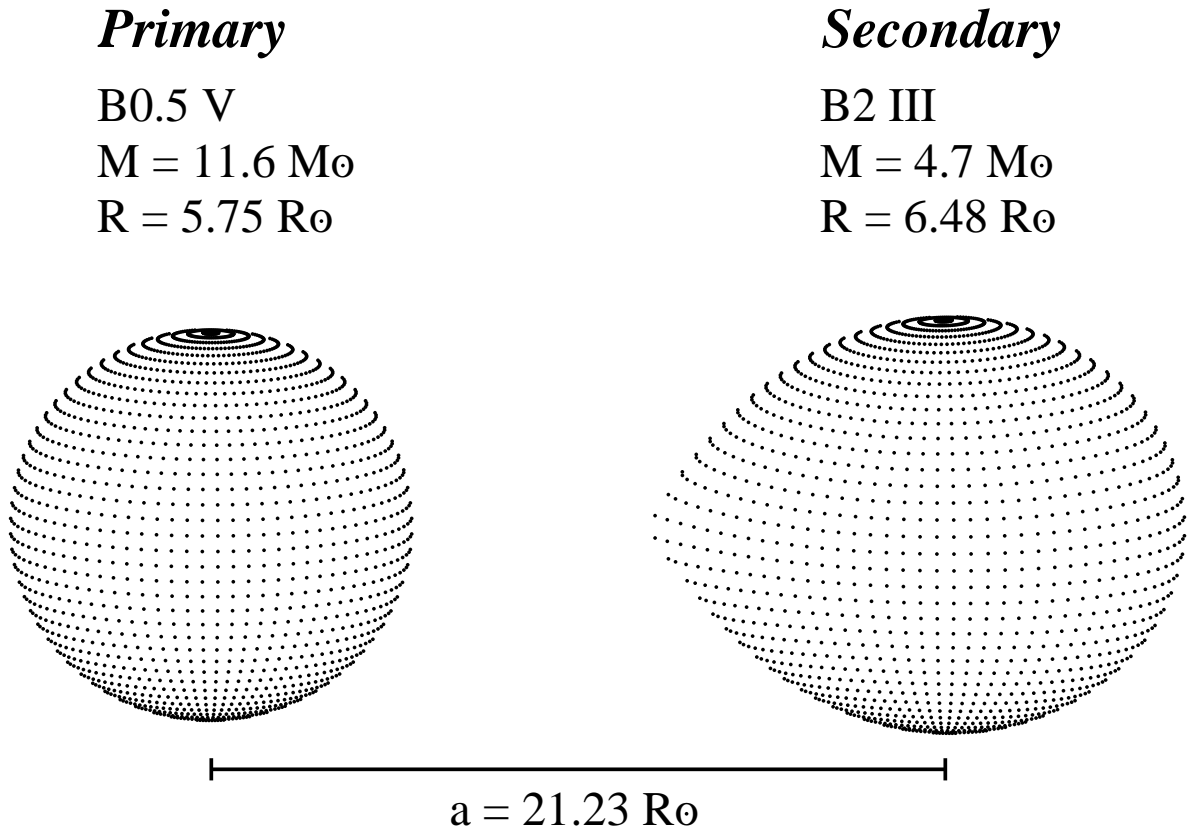}
\caption[hv5936_fig3D.eps]{A scale model of the HV~5936 system.  The sizes of the stars and their separations are shown in their correct proportions.
\label{fig3D}}
\end{figure*}

\begin{figure*}
\plotone{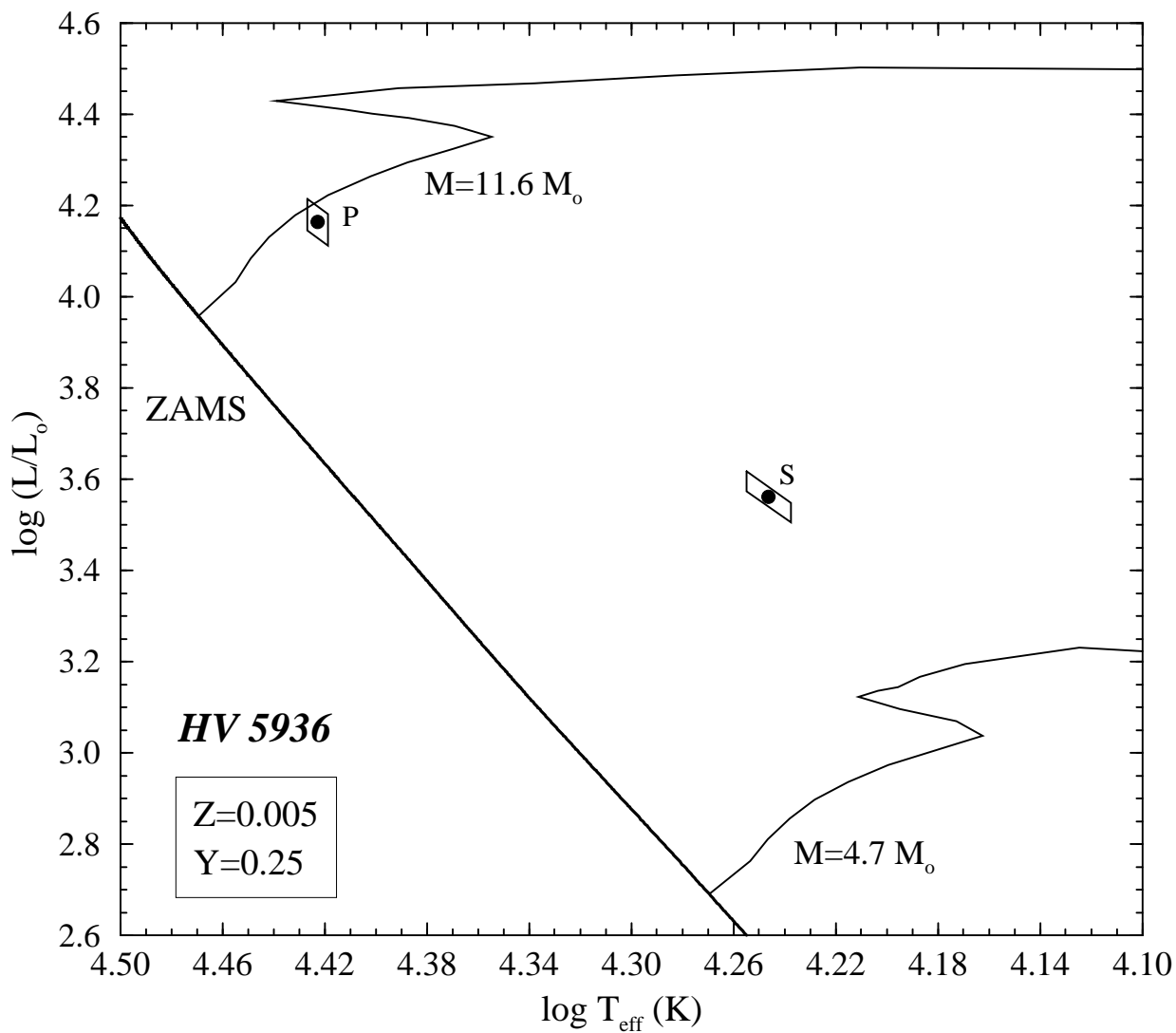}
\caption[hv5936_figHRD.eps]{A comparison of the HV~5936 results with
stellar evolution theory.  The positions of the the primary (P) and
secondary (S) components on the $\log L$ vs. $\log T_{\rm eff}$ diagram
are indicated by the filled circles.  The skewed rectangles represent
the 1$\sigma$ error boxes.  The two stellar evolution tracks shown
(solid curves) correspond to the masses derived from the binary
analysis and the metallicity measured from the UV/optical
spectrophotometry. The primary component appears to be in good
agreement with stellar evolution model predictions, while the secondary
is overluminous for its mass. This same pattern is observed in most of
the galactic Algol binaries.
\label{figHRD}}
\end{figure*}

We summarize the basic physical properties of the HV~5936 system in
Table \ref{tabSTAR}. The notes to the Table indicate how the individual
stellar properties were derived from the results of the preceding
sections.  A scale model of the system is shown in Figure \ref{fig3D}.
The locations of the HV~5936 components in the $\log T_{\rm eff}$ vs.
$\log L$ diagram are shown in Figure \ref{figHRD}, where the skewed
rectangular boxes indicate the 1$\sigma$ error loci (recall that
errors in $T_{\rm eff}$ and $L$ are correlated).  The position of the
Zero Age Main Sequence (ZAMS) is indicated in Figure 9 by the thick
curve; the theoretical evolution tracks shown on the figure will be
discussed below.

Examination of Figure \ref{figHRD} and Table \ref{tabSTAR} suggests a
paradox, in that the more evolved star (i.e., farthest from the ZAMS)
is the less massive component of the pair. The most plausible
explanation for this scenario is that HV~5936 is a post-mass transfer
system, as commonly argued to explain the so-called ``Algol paradox.''
The originally more massive star evolved beyond the Roche critical
surface at a certain point and some fraction of its mass was
transferred to its companion. After that process, the mass-accreting
component became more massive than the mass-donor, and this is the
current status of the system.

In contrast with our previous studies, the strong interaction between
the system components prevents us from using evolutionary models to
perform a critical self-consistency check of the results. I.e., a single isochrone is not expected to fit both components
of the system because of their history of mass transfer.  However,
comparison with theoretical evolution tracks is still instructive.
Thus, we considered the evolutionary models of Claret (1995, 1997) and
Claret \& Gim\'enez (1995, 1998) (altogether referred to as the CG
models). These models cover a wide range in both metallicity ($Z$) and
initial helium abundance ($Y$), incorporate the most modern input
physics, and use a value of 0.2~H$_{\rm p}$ as the convective
overshooting parameter. In our particular case, we adopted the metal
abundance from the SED fit (see Table \ref{tabSTAR}), which results in
a value of $Z=0.005$. Using this metal abundance, the empirical
chemical law of Ribas et al. (2000b) yields a helium abundance of
$n({\rm He})/n({\rm H}) = 0.084$, corresponding to $Y=0.25$. (The
analysis of the disentangled spectrum of the primary in \S 4 provides a
remarkable confirmation of this result.)

In Figure \ref{figHRD} we show the CG models for masses of
11.6~M$_{\sun}$ and 4.7~M$_{\sun}$, corresponding to the masses
measured for the HV~5936 components. Despite its history of significant
mass gain, the primary component has a luminosity in excellent
agreement with theoretical expectations for a star of its mass and
radius. This is not surprising since it has been shown that the
hydrodynamical relaxation time of the accretion process is short enough
so that the mass-gaining component in a non-conservative system is
expected to behave like a normal star of its mass and radius (De
Gr\`eve 1991, 1993). Empirical proof of this has been obtained from the
study of galactic Algol systems (Garc\'{\i}a \& Gim\'enez 1990). The
secondary, on the other hand, is significantly over-luminous for its
current mass.  This is consistent with the behavior exhibited by the
mass-donors in galactic Algol systems (Giuricin, Mardirossian, \&
Mezzetti 1983; Hilditch \& Bell 1987).

The rotational velocity determined from the disentangled spectrum for
the secondary is consistent with synchronization with
the orbit (compare the values of $v \sin i$ and $v_{syn}$ in Table
\ref{tabSTAR}).  This is expected since the secondary completely fills
its Roche lobe and tidal forces are very efficient at locking the
star's spin rate to follow the orbital motion.  This agreement between
observations and theoretical expectations provides a consistency check
on the measurements of $v \sin i$, stellar radius, and orbital
properties.  The primary component, however, is seen to have a
rotational velocity that is $\sim$25\% larger than the synchronization
value. The departure from unity is well beyond the observational
uncertainties --- but is not unprecedented or unexpected.  The
primaries of numerous galactic Algol systems have been determined to
rotate faster than their synchronous rate and even close to the
centrifugal limit (see, e.g., van Hamme \& Wilson 1990).  A rotational
velocity 25\% larger than synchronicity is well within the observed
range. Angular momentum transfer through mass accretion is probably the
most plausible model to explain the spin-up of Algol primaries (Huang
1966).

The $\sim$factor-of-2 enhancement observed in the secondary's surface
He abundance, as discussed in \S 4, provides a strong constraint on the
original masses of the stars in the HV~5936 system, particularly the
secondary.  The nature of this constraint is as follows: the initial
mass of the current secondary must have been such that, by the time the
star evolved to fill its Roche Lobe (and commence mass transfer), the
point in its interior where the He abundance was twice the surface
value was at a radius which enclosed 4.7~$M_{\sun}$ of material (i.e.,
the current mass of the secondary).  This interior point --- now at the
surface of the secondary --- clearly must have been in the outer
regions of its original H-burning core.

To exploit this constraint, we considered the evolution of 4 different
simulated binary systems, constructed with the assumptions of: 1)
non-conservative mass exchange with 50\% efficiency in the transfer of
mass from the donor star to the mass gaining star and 2) a fractional
angular momentum loss proportional to the fractional mass lost to the
system.  The initial masses of these systems were 12~$M_{\sun}$
(current secondary) + 8~$M_{\sun}$ (current primary), 14~$M_{\sun}$ +
7~$M_{\sun}$, 16~$M_{\sun}$ + 6~$M_{\sun}$, and 18~$M_{\sun}$ +
5~$M_{\sun}$.  The calculations were done following the formalism in
Torres, Neuh\"auser, \& Wichmann (1998) but considering the
non-conservative mass transfer expression in Vanbeveren et al. (1979).
From these, we estimate initial orbital periods of 1.154, 1.318, 1.689,
and 2.449 days, and initial Roche lobe radii of 4.8, 5.2, 6.1, 7.8, and
10.7~$R_{\sun}$, respectively, for the four mass combinations.  The
results showed that each of the systems experiences Case A mass
transfer (i.e., occurring during the core hydrogen burning phase of the
mass donor) and all could eventually evolve to the configuration
observed for HV~5936 --- but with different resultant surface He
abundances for the mass donor. Using internal composition profiles
computed by one of us (A.C.), we find that the surface He abundances of
the initially 12~$M_{\sun}$, 14~$M_{\sun}$, 16~$M_{\sun}$, and
18~$M_{\sun}$ mass donors are enhanced by factors of 1.5, 2.6, 3.0, and
3.5, respectively, by the time donor has been reduced to
4.7~$M_{\sun}$. The observed enhancement of $\sim$2.0 is roughly midway
between the 12~$M_{\sun}$ and 14~$M_{\sun}$ cases.

Taking into account all the results above, we postulate that the
HV~5936 system began with initial masses of $\sim$13~$M_{\sun}$ and
$\sim$7.5$~M_{\sun}$ for the current secondary and current primary,
respectively.  Case A mass exchange then resulted in the observed
current masses of the components, the spinning up of the primary to a
rotational rate faster than the synchronous rate, and the uncovering of
processed material in the atmosphere of the secondary.  The estimated
initial mass of the secondary depends only weakly on our assumption of
a 50\% mass transfer efficiency (which is probably an upper limit).
The primary's initial mass is much more sensitive to this assumption,
although it might be possible to constrain it further (and thus the
transfer efficiency) by considering the spin-up process.  Along with
the He enhancement, the secondary's surface should also exhibit a
$\sim$70$\times$ enhancement of $^{14}$N and a $\sim$10$\times$
depletion of $^{12}$C and $^{16}$O, according to our interior
composition models.  These enhancements and depletions were assumed in
the Case II modeling of the secondary's disentangled spectrum as
discussed in \S 4 above.  Because of the already low metallicity of
this LMC star, the modifications in CNO abundances do not have striking
effect on the secondary's spectrum.  In Figure \ref{figSECON} we
indicate the locations of the most prominent N II lines.  Even with
their enhanced abundances, these lines are too weak to claim a positive
detection, although there do appear to be features in the spectrum
at their locations.  Higher quality spectra will be required to confirm
their presence and, thus, the surface enhancement of N.

\section{The Interstellar Medium Towards HV~5936}

Our analysis of HV~5936 provides some insight into the conditions of
the interstellar medium along the line of sight to the system. In this
section we examine this information and also show how these data help
constrain the relative position of HV~5936 within the LMC.

The column density of interstellar H~{\sc i} in the foreground of
HV~5936 was measured by comparing the observed H~{\sc i} Ly$\alpha$
1216.7 \AA\/ absorption line profile (using the unbinned FOS data) with
theoretical profiles consisting of a synthetic stellar spectrum
convolved with an interstellar absorption profile.  In general, the
interstellar profile is constructed by assuming a component at 0 km
s$^{-1}$ with $\mbox{N(H~{\sc i})} = 5.5\times10^{20}$~cm$^{-2}$,
corresponding to Milky Way foreground gas (see, e.g., Schwering \&
Israel 1991), and a second component with a LMC-like velocity and a
column density which is varied to produce the best fit to the data (as
judged by visual inspection).

\begin{figure*}
\plotone{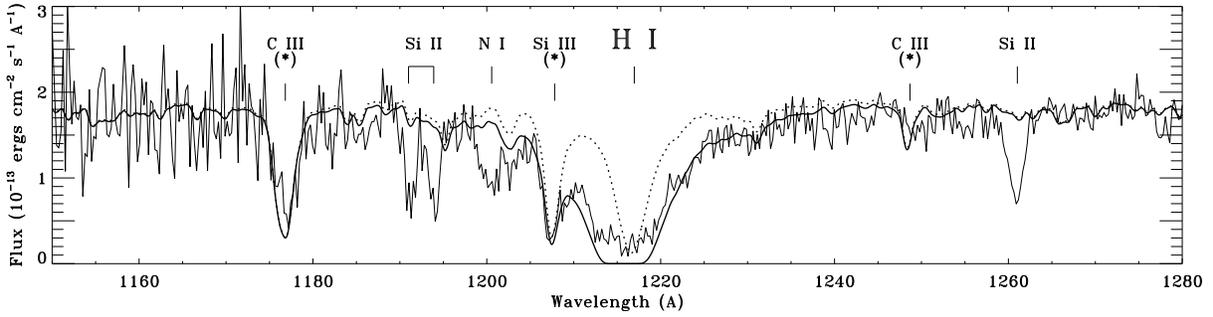}
\caption[hv5936_figHI.eps]{Derivation of the interstellar H~{\sc i}
column density towards HV~5936.  The FOS data centered on the H~{\sc i}
Ly$\alpha$ line at 1215.7 \AA\/ are shown (thin solid line).  Prominent
stellar features (denoted with an asterisk) and interstellar features
are labeled.  The dotted line represents a synthetic spectrum of the
HV~5936 system, constructed by combining two individual velocity-shifted
spectra. The individual spectra were computed using Ivan Hubeny's {\sc
synspec} spectral synthesis program with Kurucz {\sc atlas9} atmosphere
models of the appropriate stellar parameters as inputs.  The solid
curve shows the synthetic spectrum convolved with an interstellar
H~{\sc i} Ly$\alpha$ line computed with a Galactic foreground component
of $\mbox{N(H~{\sc i})} = 5.0\times10^{20}$~cm$^{-2}$ at 0 km s$^{-1}$
(see text in \S 4). \notetoeditor{THIS FIGURE IS INTENDED TO SPAN TWO
COLUMNS} 
\label{figHI}}
\end{figure*}

In examining the FOS data for HV~5936, we found no evidence for H~{\sc
i} absorption at LMC velocities, and a Milky Way contribution at 0 km
s$^{-1}$ of ${\rm 5.0\times10^{20}~cm^{-2}}$, slightly less than, but
consistent with, the mean value of Schwering \& Israel.  The
best-fitting Ly$\alpha$ profile is shown in Figure \ref{figHI}, where
we illustrate the unbinned FOS spectrum, the synthetic stellar spectrum
(dotted line) and the convolution of the synthetic spectrum with the
interstellar profile (thick solid curve).  Note that the bottom of the
Ly$\alpha$ profile is filled in by geocoronal emission.

This notable absence of neutral LMC interstellar gas towards HV~5936 is
consistent with the interstellar reddening results. The value of ${\rm
E(B-V) = 0.047\pm0.004~mag}$ found for HV~5936 from the SED analysis is
essentially identical to the mean value for 56 Milky Way foreground
stars within 1\degr\/ of the HV~5936 line-of-sight (${\rm E(B-V ) =
0.048}$ from the data of Oestreicher, Gochermann, \& Schmidt-Kaler
1995).  Thus, there is no measurable contribution to ${\rm E(B-V)}$
from LMC material.

\begin{figure*}
\plotone{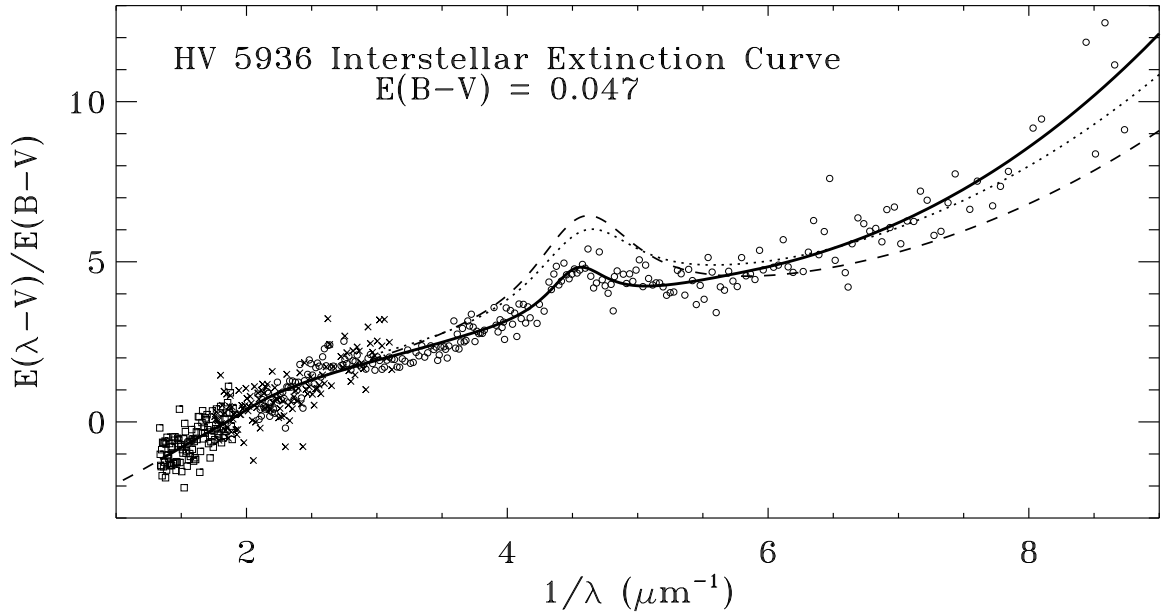}
\caption[hv5936_figEXT.eps]{Normalized UV-through-optical interstellar
extinction curve for HV~5936.  The thick solid line shows the
parameterized form of the extinction curve as determined by the SED
fitting procedure.  The recipe for constructing such a ``custom''
extinction curve is taken from F99 and the parameters defining it are
listed in Table \ref{tabEXT}.  Small symbols indicate the actual
normalized ratio of model fluxes to observed fluxes: circles, crosses,
and squares indicate FOS, STIS G430L, and STIS G750L data,
respectively.  Shown for comparison are the mean Milky Way extinction
curve for R = 3.1 from F99 (dashed line) and the mean LMC and 30
Doradus curves from Fitzpatrick (1986; dotted and dash-dotted lines,
respectively).  Based on the small value of ${\rm E(B-V)}$, the HV~5936
curve likely arises only from dust in the Milky Way halo. 
 \label{figEXT}}
\end{figure*}

The lack of large amounts of LMC gas and dust towards HV~5936 provides
an opportunity to examine some properties of the Milky Way ISM along a
sightline passing completely through the Milky Way's halo --- but free
of extragalactic contamination.  In particular, the gas-to-dust ratio
for this material is $\mbox{N(H~{\sc i})/E(B-V)} =
1\times10^{22}$~cm$^{-2}$~mag$^{-1}$, which is roughly twice the value
found in the Galactic disk and suggests a ``cleaner'' ISM in the halo.
In addition, the wavelength-dependent extinction curve derived from the
SED analysis is now recognized to be a measure of the extinction
properties of Galactic halo dust grains.  This curve is shown in Figure
\ref{figEXT} where the small symbols indicate the normalized ratio of
model fluxes to observed fluxes, while the thick solid line shows the
parameterized representation of the extinction curve, which was
actually determined by the fitting process.  The most remarkable
feature of this curve is the very weak 2175 \AA\/ bump, which may be
the weakest bump ever seen in Milky Way dust.  The curve is consistent
with the results of Kiszkurno-Koziej \& Lequeux (1987), which show a
weak trend of weakening bump strength and steepening 1500--1800 \AA\/
extinction with increasing height above the Galactic plane.  The shape
of the HV~5936 curve is very similar to those seen towards HV~982 and
EROS~1044 (Papers II and III), whose total extinctions are dominated by
Milky Way dust.

The ISM results for HV~5936 provide constraints on the position of the
system within the LMC.  Its apparent location places it within
the outlines of LMC-4, the largest supergiant H~{\sc ii} shell in the
LMC (Goudis \& Meaburn 1978; Meaburn 1980).  The shell has a diameter
of $\sim$1.1 kpc, and HV~5936 is positioned NE of the shell center and
near the inside edge.  A hint of the shell can be seen in Figure
\ref{figLMC}.  The position of the shell coincides with a ``hole'' in
the H~{\sc i} distribution (McGee \& Milton 1966).  The H~{\sc i} 21-cm
emission line data of Rohlfs et al. (1984), obtained with a half-power
beam width of 15\arcmin, reveal a total LMC H~{\sc i} column density of
${\rm 2.4 \times 10^{20} cm^{-2}}$ at the position of HV~5936, with
higher values to the east and comparable values in other directions.
The minimum value of $\mbox{N(H~{\sc i})}$ seen in the general vicinity
of HV~5936 is ${\rm 1.6\times10^{20}cm^{-2}}$, at a position 0\fdg4
west of the system.  While we have not established a rigorous upper
limit for the non-detection of LMC H~{\sc i} Ly$\alpha$ absorption
towards HV~5936, such a limit is certainly below ${\rm
10^{20}~cm^{-2}}$.  Thus, while we cannot rule out the presence of
localized, deep holes in the ISM distribution, the simplest explanation
of both the HV~5936 gas and dust results is that the system lies at an
undetermined distance above the main H~{\sc i} layer at its apparent
location on the LMC.

\section{The Distance to HV~5936}

The discussion in \S 5 demonstrates that, despite the
interesting evolutionary state of the HV~5936 system and the mass
exchange that has occurred, the stellar properties appear
well-determined and well-understood.  Particularly, the properties of
the primary star are indistinguishable from those of a ``normal'' single star,
which came by its mass through the normal star-formation process.  We
can thus determine a distance to the system based on the results for
the primary star, by combining its absolute radius $R_{\rm P}$ (derived
from the classical EB analysis) with the distance attenuation factor
$(R_{\rm P}/d)^2$ (derived from the SED analysis).  We find $d_{\rm
HV5936} = 43.2\pm 1.8$ kpc, corresponding to a distance modulus of
$(V_0 - M_V)_{\rm HV5936} = 18.18\pm0.09$ mag.

The uncertainty in the distance is estimated from considering three
independent sources of error: (1) the internal measurement errors in
$R_{\rm P}$ and $(R_{\rm P}/d)^2$ as given in Table \ref{tabPARMS}; (2)
uncertainty in the appropriate value of the extinction parameter
$R(V)$; and (3) uncertainty in the FOS flux scale zeropoint due to
calibration errors and instrument stability.  Straightforward
propagation of errors shows that these three factors yield individual
uncertainties of $\pm1.7$ kpc, $\pm0.3$ kpc (assuming $\sigma R(V) =
\pm0.3$), and $\pm0.4$ kpc (assuming $\sigma f(FOS) = \pm2.5$\%),
respectively.  The overall 1$\sigma$ uncertainty quoted above is the
quadratic sum of these three errors.  It is dominated by the
uncertainty in the primary's radius, which alone accounts for 1.7 kpc
in the error budget. Note that the only ``adjustable'' factor in the
analysis is the extinction parameter $R(V)$, for which we have assumed
the value 3.1.  Because of the very low reddening of the system, our
distance result is very insensitive to this assumption. The weak
dependence of the distance modulus on $R(V)$ is given by: $(V_0 -
M_V)_{\rm HV5936} = 18.18 - 0.04 \times [R(V)-3.1]$.

We have considered the possibility that this distance result may be
tainted due to the advanced evolutionary state of the secondary star.
In general, the light curve and radial velocity curve analyses are very
insensitive to the natures of the EB's components, i.e., the
temperature ratio of the two stars, their mean radii, surface
gravities, and relative contributions to the observed light in the
optical spectral region are virtually model-independent and very
well-determined.  These results tightly constrain the SED analysis and
so it would seem that the results could be compromised only if the
secondary's SED departs significantly from the shape predicted by an
{\sc atlas9} model of the derived $T_{\rm eff}$, $\log g$, [m/H], and
$v_{\rm micro}$ values. The evidence, however, suggests that this is
not the case.  In support, we note the high quality of the fit to the
combined SED of the system, which is comparable to our previous studies
of systems containing wholly ``normal'' stars, and the consistency
between the SED analysis and the independent modeling of the
disentangled optical spectra (see \S 4).  The only anomaly found is the
probable modest enhancement of the secondary's He abundance, which
would have only very small effect on the star's SED.  We have run a
number of tests with the SED analysis, arbitrarily modifying the
secondary's SED (by varying $T_{\rm eff}$, $\log g$, [m/H], or $v_{\rm
micro}$).  Even relatively large changes have little effect on
the derived system distance, because of the dominance of the primary in
the combined SED.  We thus conclude, based both on a lack of contrary
evidence and on the apparent robustness of the result, that
uncertainties arising from the evolutionary state of the secondary are
likely significantly smaller than the other sources of error noted
above.

In Table \ref{tabDIST} we show the individual system distances for
HV~5936 and the other EBs in our study.  We also list the implied
distances to a standard reference point in the LMC, based on the
assumption that the EBs are all located within a flat disk-like LMC
whose spatial orientation is specified by an inclination angle and a
position angle for the line of nodes in the plane of the sky.  We
choose the optical center of the LMC's Bar as the reference point and
show two sets of LMC distances, based on two different assumptions
about the LMC's spatial orientation.  Details are given in the Notes to
the Table.

The results for HV~5936 in Table \ref{tabDIST} clearly stand out from
the others.  This is partially explained by projection effects.  Because
HV~5936 lies relatively far from the adopted reference point and in the
``nearside'' of the LMC, it is expected to be somewhat closer than the
other systems.  However, when the projection is taken into account, the
distance implied for the LMC reference point is $44.3\pm1.8$ kpc,
corresponding to a distance modulus of $18.23\pm0.09$ (from taking a
simple mean of the two cases shown in Table \ref{tabDIST}).  This is
about 2$\sigma$ away from the mean of the other systems and suggests
that the assumption that all the systems lie in the same flat disk may
be invalid.

Although the relatively large uncertainty in the HV~5936 result
prevents definitive conclusions, we suggest that HV~5936 lies ``above''
the extrapolated position of the LMC's disk.  This notion is consistent
with, and actually suggested by, the interstellar results, which reveal
no evidence for absorption by LMC H~{\sc i} and no measurable
extinction by LMC dust along the HV~5936 sightline.  This interpretation
of the ISM data is complicated by the coincidence of HV~5936's position
in the sky with the supergiant shell LMC~4, towards which a very low
H~{\sc i} 21-cm emission column density is observed.  Depending on the
patchiness of the H~{\sc i}, it is conceivable that a star located
within or even behind LMC~4 could show an undetectably low H~{\sc i}
absorption column density.  This seems unlikely for HV~5936, however,
since it is located near the eastern edge of the LMC~4, where the 21-cm
column density is ${\rm \sim 2.4\times 10^{20}~cm^{-2}}$ and rises
steeply towards the east.  The simplest explanation is
that HV~5936 just lies in front of most of the LMC's interstellar
matter.  It is not clear why HV~5936 should occupy such a position,
although this might well be related to the formation of shell LMC~4
itself.  Likewise, it remains to be seen whether HV~5936 is a
pathological object, in terms of its location, or is merely
representative of a spatially extended stellar population in the NE
quadrant of the LMC.

Alternative explanations for the HV~5936 results include assuming
simple error in the spatial orientation of the LMC or perhaps a major
structural feature, such as a warp, in the LMC's disk.  The latter may
arise from strong past interactions between the Magellanic Clouds and
the Milky Way galaxy. Both hypotheses would allow HV~5936 to lie close
to the LMC's disk (but still above, given the ISM results), but require
the disk to be closer to us than currently assumed.  Unfortunately for
these ideas, the required orientation for the LMC would necessitate an
inclination angle ($\sim60$\arcdeg) much steeper than previolusly
measured.  Moreover, structural studies of the LMC have revealed no
evidence of a warp in the vicinity of HV~5936 (e.g., see Olsen \& Salyk
2002).
  
\section{Final Comments}

The analysis presented here for HV~5936 has differed somewhat from
those reported previously in Papers I, II, and III.  Here our efforts
have been focused more strongly on the physical properties of the
stellar components and the evolutionary history of the binary system,
rather than on the distance measurement.  This arises from the
interesting evolutionary state of this semidetached system and from its
outlying location in the LMC. Because of its location, the implications of
HV~5936's individual distance for the distance to the LMC as a whole is
not as firmly established as for our previously analyzed systems.

It is worth noting that HV~5936 is the first semidetached EB system to
be used as a distance indicator for the LMC. Although our current
analysis shows that semidetached binaries can be used successfully as
such, there are a number of complications that call for the careful use
of these systems. Among the related problems are: {\em 1)} The lack of
evolutionary cross checks, {\em 2)} the often very unequal component
masses, effective temperatures and luminosities that complicate the
spectrophotometric analysis, {\em 3)} the large out-of-eclipse
variations and distortions, and {\em 4)} the relative radii that are
strongly dependent on the adopted mass ratio.  Our success with HV~5936
springs directly from the high quality observations available, in which
the spectroscopic and photometric signatures of both stars are clearly
present and separable.

Future distance studies within our program will be focused primarily on
non-interacting systems lying closer to the apparent center of the
LMC.  We currently have four detached systems under analysis, all of
which can be expected to yield accurate distances and a number of cross
checks to verify the results. In addition, two of these systems, EROS
1066 and MACHO 053648.7$-$691700 (see Figure 1), have been especially
selected to provide insight into the possible problem of line-of-sight
extension of the LMC.

\acknowledgements

This work was supported by NASA grants NAG5-7113, HST GO-06683, HST
GO-08691, and NSF/RUI AST-0071260.  We are grateful for the skilled
assistance of the CTIO support staff during our observing runs.  E.F.
acknowledges support from NASA ADP grant NAG5-7117 to Villanova
University and thanks Michael Oestreicher for kindly making his LMC
foreground extinction data available. 


\clearpage
  

\begin{deluxetable}{ccccccc}
\tablewidth{0pc}
\tablecaption{HST Spectrophotometric Observations of HV~5936}
\tablehead{
\colhead{}             & 
\colhead{}             & 
\colhead{Dataset}      & 
\colhead{Date of}      & 
\colhead{Orbital}      & 
\colhead{ }            & 
\colhead{ }           \\ 
\colhead{Instrument}             & 
\colhead{Detector}               & 
\colhead{Name}                   & 
\colhead{Observation}            & 
\colhead{Phase}                  & 
\colhead{$k_P$\tablenotemark{a} }  & 
\colhead{$k_S$\tablenotemark{a} }  }
\startdata
FOS  & G130H & Y3FU0803T & 31 Jan 1997   & 0.115 & 1.0059 & 0.9875 \\
FOS  & G190H & Y3FU0806T & 31 Jan 1997   & 0.137 & 1.0116 & 1.0151 \\
FOS  & G270H & Y3FU0805T & 31 Jan 1997   & 0.133 & 1.0106 & 1.0101 \\
FOS  & G400H & Y3FU0804T & 31 Jan 1997   & 0.128 & 1.0093 & 1.0038 \\
STIS & G430L & O665B6030 & 22 April 2001 & 0.654 & 1.0491 & 1.0360 \\
STIS & G750L & O665B6040 & 22 April 2001 & 0.657 & 1.0490 & 1.0396 \\   
\tablenotetext{a}{The quantities $k_P$ and $k_S$ are corrections used in the spectral energy distribution analysis to account for out-of-eclipse light variations in the HV~5936 system, due to reflection effects and gravitational distortion of the secondary. See \S 3.3.} 
\enddata
\label{tabHST}
\end{deluxetable}
\clearpage


\begin{deluxetable}{rrrrrr}
\tablewidth{0pc}
\tablecaption{Heliocentric Radial Velocity Measurements for HV~5936}
\tablehead{
\colhead{HJD}              & 
\colhead{Orbital}          &  
\colhead{RV$_{\rm P}$}     &
\colhead{RV$_{\rm S}$}     &
\colhead{(O--C)$_{\rm P}$} &
\colhead{(O--C)$_{\rm S}$} \\
\colhead{($-$2400000)}  &
\colhead{Phase}         & 
\colhead{(km s$^{-1}$)} &
\colhead{(km s$^{-1}$)} &
\colhead{(km s$^{-1}$)} &
\colhead{(km s$^{-1}$)} }
\startdata
51558.7339 & 0.6719 & 411.2 &  84.6 &    1.8 &    2.0\\
51558.7600 & 0.6812 & 413.7 &  77.5 &    1.5 &    1.3\\
51560.7402 & 0.3871 & 240.4 & 486.6 & $-$1.9 & $-$6.2\\
51560.7659 & 0.3963 & 246.8 & 475.4 & $-$0.3 & $-$5.8\\
51561.6849 & 0.7239 & 421.1 &  55.8 & $-$0.2 & $-$1.1\\
51561.7072 & 0.7318 & 421.6 &  53.5 & $-$0.5 & $-$1.6\\
51562.7743 & 0.1122 & 245.7 & 484.5 &    1.3 &    1.0\\
51562.7930 & 0.1189 & 240.6 & 490.5 & $-$0.3 & $-$1.2\\
51895.7673 & 0.8234 & 412.8 &  76.3 &    0.2 & $-$1.3\\
51895.7906 & 0.8317 & 408.4 &  81.2 & $-$1.6 & $-$2.4\\
51898.6384 & 0.8470 & 402.7 &  96.0 & $-$1.9 &    0.0\\
51898.6616 & 0.8553 & 402.7 & 102.1 &    1.3 & $-$1.6\\
51899.6325 & 0.2014 & 211.9 & 565.6 &    1.1 &    2.8\\
51899.7847 & 0.2556 & 206.5 & 578.1 &    1.0 &    3.0\\
51900.8064 & 0.6199 & 387.3 & 133.8 & $-$0.3 &    1.3\\
51900.8296 & 0.6281 & 391.9 & 127.4 &    0.2 &    4.5\\
51902.6812 & 0.2882 & 207.6 & 569.7 & $-$0.7 &    0.4\\
51902.7046 & 0.2966 & 209.6 & 568.0 & $-$0.2 &    1.8\\
\enddata
\label{tabRV}
\end{deluxetable}
\clearpage


\small
\begin{deluxetable}{lc}
\tablewidth{0pc}
\tablecaption{Results From Light Curve, Radial Velocity Curve, and Spectrophotometry Analyses}
\tablehead{\colhead{Parameter} & \colhead{Value}}
\startdata
\multicolumn{2}{c}{{\it Light Curve Analysis}}    \\
Period                                                & $2.8050681\pm0.0000015$ days \\
Eccentricity                                          & $0.0$ (fixed)         \\
Inclination                                           & $80.0\pm0.2$ deg      \\
$T_{\rm eff}^{\rm S}/T_{\rm eff}^{\rm P}$             & $0.666\pm0.008$       \\
$[L_{\rm S}/L_{\rm P}]_{\rm B}$   		      & $0.592\pm0.011$       \\
$[L_{\rm S}/L_{\rm P}]_{\rm V}$   		      & $0.629\pm0.012$       \\
$r_{\rm P}$\tablenotemark{a}                          & $0.2708\pm0.0088$     \\
$r_{\rm S}$\tablenotemark{a}                          & $0.3053\pm0.0034$     \\
$\Omega_P$\tablenotemark{b}                           & $4.14\pm0.15$         \\
$\Omega_S$\tablenotemark{b}                           & $2.69\pm0.04$         \\  
$\Delta\phi$                                          & $0.0006\pm0.0005$      \\

\multicolumn{2}{c}{{\it Radial Velocity Curve Analysis}}                         \\
$K_{\rm P}$                                           & $109.1\pm3.2$ km~s$^{-1}$\\
$K_{\rm S}$                                           & $268.2\pm7.5$ km~s$^{-1}$\\
$q$\tablenotemark{e}                                  & $0.407\pm0.016$          \\
$v_{\gamma}$                                          & $314.3\pm5.8$ km~s$^{-1}$\\
$a$                                                   & $21.23\pm0.46$ R$_{\odot}$\\
\multicolumn{2}{c}{\it Energy Distribution Analysis}                             \\
$T_{\rm eff}^{\rm P}$                                 &  $26,450\pm250$ K      \\
${\rm[m/H]}_{\rm PS}$                                 &  $-0.63\pm0.05$\phm{$-$} \\
$v_{\rm micro}^{\rm PS}$                              &  $2.6\pm0.6$ km~s$^{-1}$ \\      
E(B$-$V)                                              &  $0.047\pm0.005$ mag  \\
$\log (R_P/d)^2$                                      &  $-23.046\pm0.006$    \\
\tablenotetext{a}{Fractional stellar radius $r \equiv R/a$, where 
$R$ is the stellar ``volume radius'' and $a$ is the orbital semi-major axis.}
\tablenotetext{b}{Normalized potential at stellar surface.}
\tablenotetext{c}{Mass ratio ${M_{\rm S}}/{M_{\rm P}}$.}
\enddata
\normalsize
\label{tabPARMS}
\end{deluxetable}
\normalsize
\clearpage


\begin{deluxetable}{ccc}
\tablewidth{0pc}
\tablecaption{Offsets Applied to HST/STIS Observations of HV~5936}
\tablehead{
\colhead{STIS}  & 
\colhead{Dataset} & 
\colhead{Offset}       \\
\colhead{Detector}         & 
\colhead{Name}            & 
\colhead{(${\rm FOS - STIS}$)} }
\startdata
 G430L & O665B6030  &  +$10.2\pm0.5\%$  \\ 
 G750L & O665B6040  &  \phn+$5.9\pm0.7$\%\\ 
\enddata
\label{tabSTIS}
\end{deluxetable}
\clearpage


\begin{deluxetable}{clc}
\tablewidth{0pc}
\tablecaption{Extinction Curve Parameters for HV~5936}
\tablehead{
\colhead{Parameter}   & 
\colhead{Description} & 
\colhead{Value}        }
\startdata
$x_0$       & UV bump centroid        & $4.55\pm0.03$ ${\rm\mu m^{-1}}$\\
$\gamma$    & UV bump FWHM            & $0.59\pm0.18$ ${\rm\mu m^{-1}}$\\
$c_1$       & linear offset           & $-1.01\pm0.40$                 \\
$c_2$       & linear slope            & $\phm{-}1.02\pm 0.10$          \\
$c_3$       & UV bump strength        & $\phm{-}0.53\pm 0.26$          \\
$c_4$       & FUV curvature           & $\phm{-}0.61\pm 0.12$          \\   
$R(V)$         & ${\rm A(V)/E(B-V)}    $  & 3.1 (assumed)               \\
\enddata
\tablecomments{The extinction curve parametrization scheme is based on the work of Fitzpatrick \& Massa 1990 and the complete UV-through-IR curve is constructed following the recipe of Fitzpatrick 1999.}
\label{tabEXT}
\end{deluxetable}
\clearpage

 
\begin{deluxetable}{lccc}
\tablewidth{0pc}
\tablecaption{Analysis of Disentangled Spectra of HV~5936 Components}
\tablehead{
\colhead{Stellar Property}      & 
\colhead{Primary}               &  
\colhead{Secondary}             &  
\colhead{Secondary}             \\
\colhead{  }                    & 
\colhead{  }                    &  
\colhead{Case I}                &  
\colhead{Case II}               \\
\colhead{(1)}                   &
\colhead{(2)}                   &  
\colhead{(3)}                   &  
\colhead{(4)}                   } 
\startdata
n(He)/n(H)                      & 0.084\tablenotemark{a} & 0.084\tablenotemark{a}     & 0.16\tablenotemark{b}    \\          
T$_{\rm eff}$ (K)               & $26,900\pm370$	 & $21,040\pm590$            & $17,620\pm250$           \\
$\log g$ (cgs)                  & $4.11\pm0.04$          & $3.73\pm0.10$              & $3.54\pm0.06$             \\
${\rm [m/H]}$                   & $-0.60\pm0.04$         & $-0.63$\tablenotemark{c}   & $-0.63$\tablenotemark{c}  \\
$v_{\rm micro}$ (km s$^{-1}$)   & 2.6\tablenotemark{d}   & 2.6\tablenotemark{d}       & 2.6\tablenotemark{d}      \\ 
$v \sin i$ (km s$^{-1}$)      	& $127.8\pm1.6$      	 & $139.0\pm5.4$              & $127.0\pm2.2$             \\
$v_{\rm radial}$ (km s$^{-1}$)         & $314.5\pm1.2$     	 & $318.0\pm1.8$              & $316.7\pm1.7$             \\
contribution to light (\%)	& $62.3\pm1.3$\tablenotemark{e}           & $40.5\pm1.2$\tablenotemark{e}               & $36.7\pm0.4$\tablenotemark{e}              \\
\tablenotetext{a}{Assumed, from the discussion in \S 5; corresponds to Y = 0.25.}
\tablenotetext{b}{Enhanced He abundance. Also the abundances of C, N, and O, were adjusted by factors of 0.1, 70, and 0.1, respectively relative to the base metallicity of $[{\rm m/H}] = -0.63$.  See the discussions in \S 4 and 5.}
\tablenotetext{c}{There are insufficient spectral features to measure [m/H] from the optical spectrum of the secondary. The value of [m/H] derived from the SED analysis in \S 3.3 is assumed.}
\tablenotetext{d}{The microturbulence velocity cannot be measured with the existing data.  The value derived from the SED analysis in \S 3.3 is assumed.}
\tablenotetext{e}{Contribution factors for the primary and secondary stars were determined independently of each other.  They were not constrained to add up to 100\%.}
\enddata
\label{tabFINE}
\end{deluxetable}

 
\begin{deluxetable}{lcc}
\tabletypesize{\footnotesize}
\tablewidth{0pc}
\tablecaption{Physical Properties of the HV~5936 System}
\tablehead{
\colhead{Property}     & 
\colhead{Primary}      &  
\colhead{Secondary}    \\
\colhead{}             &
\colhead{Star}         &  
\colhead{Star}          } 
\startdata
Spectral Type\tablenotemark{a}                       & B0.5 V                  & B2 III                        \\
Mass\tablenotemark{b} (M$_{\sun}$)                   & $11.6\pm0.5$            & $4.7\pm0.2$                  \\
Radius\tablenotemark{c} (R$_{\sun}$)                 & $5.75\pm0.23$           & $6.48\pm0.16$                 \\
$\log g$\tablenotemark{d} (cgs)                      & $3.984\pm0.039$         & $3.488\pm0.029$               \\
$T_{\rm eff}$\tablenotemark{e} (K)                   & $26,450\pm250$         & $17,600\pm330$              \\
$\log (L/L_{\sun})$\tablenotemark{f}                 & $3.98\pm0.04$           & $3.49\pm0.03$                 \\
$[$m/H$]$\tablenotemark{g}                           & $-0.62\pm0.05$          &$-0.62\pm0.05$\phm{$-$}       \\
$n({\rm He})/n({\rm H})$\tablenotemark{h}            & 0.084                   & 0.16                          \\
$v \sin i$\tablenotemark{i} (km s$^{-1}$)            & $127.8\pm1.6$           & $127.0\pm2.2$                \\
$v_{\rm sync} \sin i$\tablenotemark{j} (km s$^{-1}$) & $103.1\pm4.1$           & $123.9\pm2.9$                \\ 
$d_{\rm HV5936}$\tablenotemark{k} (kpc)              & \multicolumn{2}{c}{$43.2\pm1.8$}                      \\

\tablenotetext{a}{Estimated from $T_{\rm eff}$ and $\log g$} 
\tablenotetext{b}{From the mass ratio $q$ and the application of Kepler's Third Law.}
\tablenotetext{c}{Computed from the relative radii $r_P$ and $r_S$ and the orbital semimajor axis $a$.}
\tablenotetext{d}{Computed from $g = G M / R^2$.}
\tablenotetext{e}{Direct result of the spectrophotometry analysis and photometrically-determined temperature ratio.}
\tablenotetext{f}{Computed from $L = 4 \pi R^2 \sigma T^4_{\rm eff}$.} 
\tablenotetext{g}{Mean result from the SED analysis in \S 3.3 and the synthetic spectrum analysis for the primary in \S 5. For the secondary, abundance anomalies in CNO abundances would be expected, given the observed He enhancement.}
\tablenotetext{h}{For the primary, this corresponds to a He mass fraction of $Y = 0.25$ and is based on the observed metallicity and standard chemical enrichment. See \S 5. For the secondary, $n({\rm He})/n({\rm H})$ is based on analysis of optical spectra in \S 4.  The He enhancement is likely accompanied by modifications in the surface abundances of CNO.  See \S 5.}
\tablenotetext{i}{$v \sin i$ measured from the ``disentangled spectra'' of the two components as described in the text in \S 5.}
\tablenotetext{j}{Theoretical synchronization velocities.}
\tablenotetext{k}{Using $(R_P/d)^2$ from the spectrophotometry analysis and $R_P$ from the light curve and radial velocity curve analyses. See \S 7.} 
\enddata
\label{tabSTAR}
\end{deluxetable}
\clearpage


\small
\begin{deluxetable}{llccc}
\tablewidth{0pc}
\tablecaption{Distances to LMC EB Systems}
\tablehead{
\colhead{}           &
\colhead{}           &
\colhead{}           &
\colhead{$\rm d_{LMC}$}      &
\colhead{$\rm d_{LMC}$}      \\
\colhead{EB System}   &
\colhead{Reference}   &
\colhead{$\rm d_{EB}$} &
\colhead{(Case I)\tablenotemark{a}} &
\colhead{(Case II)\tablenotemark{b}}}
\startdata
HV~2274    & Paper I,II & $47.0\pm2.2$ kpc & $45.9$ kpc & $47.0$ kpc \\
HV~982     & Paper II   & $50.2\pm1.2$ kpc & $50.6$ kpc & $50.7$ kpc \\
EROS~1044  & Paper III  & $47.5\pm1.8$ kpc & $47.3$ kpc & $47.4$ kpc \\
HV~5936    & This Paper & $43.2\pm1.8$ kpc & $44.0$ kpc & $44.7$ kpc \\
\enddata
\tablenotetext{a}{Distance at a reference point at ($\alpha$,
$\delta$)$_{1950}$ = ($5^h 24^m$, $-69\arcdeg\/ 47\arcmin$),
corresponding to the optical center of the LMC's bar according to
Isserstedt 1975.  Adopted LMC orientation defined by an inclination
angle of $38\arcdeg$ and a line-of-nodes position angle of
$168\arcdeg$, from Schmidt-Kaler \& Gochermann 1992.  This orientation
is illustrated in Figure \ref{figLMC}.}
\tablenotetext{b}{Distance referred to the same reference point as
above.  Adopted LMC orientation defined by an inclination angle of
$34\fdg7$ and a line-of-nodes position angle of $122\fdg5$, from van
der Marel \& Cioni 2001. In this case, the line of nodes runs approximately
parallel to the long axis of the LMC's Bar.}

\label{tabDIST}
\end{deluxetable}
\clearpage


\clearpage

\clearpage

\clearpage

\clearpage

\clearpage
\small
\normalsize

\clearpage

\clearpage

\end{document}